\documentclass[twocolumn,showpacs,preprintnumbers,amsmath,amssymb,floatfix]{revtex4}
\usepackage{amsmath,amsfonts,amssymb,amsthm,epsfig,epstopdf,url,array}
\usepackage{graphicx}
\usepackage{dcolumn}
\usepackage{bm}

\newcommand{\beq}{\begin{equation}}
\newcommand{\eeq}{\end{equation}}
\newcommand{\bey}{\begin{eqnarray}}
\newcommand{\eey}{\end{eqnarray}}

\begin{document}

\title{ Compact stars in low-mass X-ray binaries}

\author{Sk. Monowar Hossein}
\email{hossein@iucaa.ernet.in} \affiliation{Department of
Mathematics, Aliah University, Sector - V , Salt Lake,  Kolkata -
700091, India}

\author{Sajahan Molla}
\email{sajahan.phy@gmail.com} \affiliation{Department of Physics,
Aliah University, Sector - V , Salt Lake,  Kolkata, India}

\author{Md. Abdul Kayum Jafry}
\email{akjafry@yahoo.com} \affiliation{Department of
Physics, Shibpur Dinobundhoo Institution (College), Howrah 711102, West Bengal,
India}

\author{ Mehedi Kalam}
\email{kalam@iucaa.ernet.in} \affiliation{Department of
Physics, Aliah University, Sector - V , Salt Lake,  Kolkata -
700091, India}

\date{\today}

\begin{abstract}
We propose a model for compact stars in low-mass X-ray binaries(LMXBs) namely KS 1731-260, EXO 1745-248 and 4U 1608-52\citep{Guver2010,Ozel2012a,Ozel2009b}. Here we investigate the physical phenomena of compact stars in the LMXBs. Using our model, we have calculated central density ($\rho _{0}$), surface density($\rho _{b}$), mass(M) and red-shift($z_{s}$) for the above mentioned compact stars, which is very much consistent with the reported data. We also obtain the possible equation of state(EOS) of the stars which is physically acceptable.
\end{abstract}

\maketitle
\section{Introduction}
Compact objects  in the low-mass X-ray binaries(LMXBs) takes much attention to the astrophysicists for the last few years. In LMXBs, possible compact objects are Neutron Star/Strange Star. Neutron stars are composed almost entirely of neutrons ,while strange stars are of quark matter, or strange matter. Few hundred of LMXBs have been detected in the Milky Way, and only very few LMXBs have been discovered in globular clusters. Chandra X-ray Observatory data has revealed LMXBs in many distant galaxies. A low-mass X-ray binary emits radiation mostly in X-rays. They emit radiation which is less than one percent in visible region. Accretion disk around the compact object is the brightest one of it. LMXBs have the orbital period range from ten minutes to several hundred days.
Among the LMXBs, the possible existence of Neutron Star/Strange Star are EXO 1745-248, KS 1731-260, 4U 1608-52\citep{Guver2010,Ozel2012a,Ozel2009b}. Many researchers studied\citep{Rahaman2012a,Kalam2012a,Hossein2012,Rahaman2012b,Kalam2012b,Kalam2013,Lobo2006,Bronnikov2006,Egeland2007,Dymnikova2002}
  compact stars in various directions. Scientists used different techniques such as computational, observational or theoretical analysis to study  astrophysical objects.Tolman\citep{Tolman1939} proposed  an isotropic static solutions for a sphere of fluid. He pointed out that due to some complexity of the VII-th solution (among the eight different solutions), it is not possible to explain the physical behavior(obviously there is a misprint in the original article in Tolman VII solution)of a star. We have some curiosities about the conclusion. Therefore, considering the isotropic model (which Tolman assume), we wanted to explain the physical behaviour of the compact objects in the low-mass X-ray binaries by taking the corrected metric.

In this article, we compare  our results ( mass, radius, central density, surface density and
 surface red-shift ) with  the compact stars  in the LMXBs and it is found to be consistent
 with the reported datas\citep{Guver2010,Ozel2012a,Ozel2009b}.\\
 We organize the article as follows:\\
  In Sec II, we have discuss the interior spacetime and behavior of the star. In Sec. III,
 we discuss the matching conditions. In Sec. IV, we have studied some special features of the star namely, TOV equation, Energy conditions, Stability,
 Mass-radius relation, Compactness, Surface redshift and possible Equation of State(EOS) in different sub-sections. The article concluded
with a short discussion with numerical datas.

\section{Interior Spacetime of the star}
We consider the interior space-time as:
\begin{eqnarray}
ds^2 = -B^2\sin ^{2}\ln \sqrt{\frac{\sqrt{1-\frac{r^2}{R^2}+4\frac{r^4}{A^4}}+2\frac{r^2}{A^2}-\frac{1}{4}\frac{A^2}{R^2}}{C}}dt^2 \nonumber \\
+\left(1-\frac{r^2}{R^2}+4\frac{r^4}{A^4}\right)^{-1}dr^2 +r^2d\Omega^{2}. ~~~~~~~~~~\label{eq1}
\end{eqnarray}
Tolman \citep{Tolman1939} proposed such type of metric (1) to develop a viable model for a spherical static isotropic fluid where $R$,$C$,$A$ and $B$ are constants.
We assume that the energy-momentum tensor for the interior of the compact star has the standard form
\begin{equation}
               T_\nu^\mu=  ( -\rho , p, p, p)
         \label{Eq2}
          \end{equation}

where $\rho$ and $p$ are energy-density and isotropic pressure respectively.

Einstein's field equations for the metric (1) accordingly are obtained as ($ c=1,G=1$)
\begin{eqnarray}
8\pi  \rho &=& \left(1-\frac{r^2}{R^2}+4\frac{r^4}{A^4}\right)\left[\frac{\lambda^\prime}{r}-\frac{1}{r^2}\right]+\frac{1}{r^2},\label{eq2}\\
8\pi  p &=& \left(1-\frac{r^2}{R^2}+4\frac{r^4}{A^4}\right)\left[\frac{\nu^\prime}{r}+\frac{1}{r^2}\right]-\frac{1}{r^2}.\label{eq3}
\end{eqnarray}

 Where
\begin{eqnarray}
\lambda^\prime &=&
\frac{\left(2\frac{r}{R^2}-16\frac{r^3}{A^4}\right)}{\left(1-\frac{r^2}{R^2}+4\frac{r^4}{A^4}\right)},
\end{eqnarray}
\begin{eqnarray}
\nu^\prime&=&\frac{\left(1-\frac{r^2}{R^2}+4\frac{r^4}{A^4}\right)^{-1/2}\left(-\frac{r}{2R^2}+4\frac{r^3}{A^4}\right)+2\frac{r}{A^2}}
{\left(\sqrt{1-\frac{r^2}{R^2}+4\frac{r^4}{A^4}}+2\frac{r^2}{A^2}-\frac{1}{4}\frac{A^2}{R^2}\right)}\nonumber\\
&&2\cot\ln\sqrt{\frac{\sqrt{1-\frac{r^2}{R^2}+4\frac{r^4}{A^4}}+2\frac{r^2}{A^2}-\frac{1}{4}\frac{A^2}{R^2}}{C}},\label{eq4}
\end{eqnarray}
\begin{eqnarray}
\nu^{\prime\prime}&=&[(\sqrt{1-\frac{r^2}{R^2}+4\frac{r^4}{A^4}}+2\frac{r^2}{A^2}-\frac{1}{4}\frac{A^2}{R^2})^{-1}\nonumber
\\&&[-2\left(1-\frac{r^2}{R^2}+4\frac{r^4}{A^4}\right)^{-3/2}\left(-\frac{r}{2R^2}+4\frac{r^3}{A^4}\right)^{2}\nonumber
\\&&+\left(1-\frac{r^2}{R^2}+4\frac{r^4}{A^4}\right)^{-1/2}\left(-\frac{1}{2R^2}+\frac{12r^2}{A^4}\right)+\frac{2}{A^2}]\nonumber
\\&&-2(\sqrt{1-\frac{r^2}{R^2}+4\frac{r^4}{A^4}}+2\frac{r^2}{A^2}-\frac{1}{4}\frac{A^2}{R^2})^{-2}\nonumber
\\&&[\left(1-\frac{r^2}{R^2}+4\frac{r^4}{A^4}\right)^{-1/2}\left(-\frac{r}{2R^2}+4\frac{r^3}{A^4}\right)+\frac{2r}{A^{2}}]^{2}]\nonumber
\\&&2\cot\ln\sqrt{\frac{\sqrt{1-\frac{r^2}{R^2}+4\frac{r^4}{A^4}}+2\frac{r^2}{A^2}-\frac{1}{4}\frac{A^2}{R^2}}{C}}\nonumber
\\&&-2[\frac{\left(1-\frac{r^2}{R^2}+4\frac{r^4}{A^4}\right)^{-1/2}\left(-\frac{r}{2R^2}+4\frac{r^3}{A^4}\right)+\frac{2r}{A^{2}}}
{\sqrt{1-\frac{r^2}{R^2}+4\frac{r^4}{A^4}}+2\frac{r^2}{A^2}-\frac{1}{4}\frac{A^2}{R^2}}\nonumber
\\&&cosec\ln\sqrt{\frac{\sqrt{1-\frac{r^2}{R^2}+4\frac{r^4}{A^4}}+2\frac{r^2}{A^2}-\frac{1}{4}\frac{A^2}{R^2}}{C}}]^{2}.
\end{eqnarray}

$\bf{Density~and~Pressure~of~the~compact~star~:}$\\

Now from eqn.(3) and eqn.(4) we have,

\begin{eqnarray}
\rho &=& \frac{1}{8\pi}\left(\frac{3}{R^2}-20\frac{r^2}{A^4}\right) ,\nonumber\\
\rho_0 &=& \frac{3}{8\pi  R^2}  ,\nonumber\\
\rho_b &=& \frac{1}{8\pi}\left(\frac{3}{R^2}-20\frac{b^2}{A^4}\right),\nonumber
\end{eqnarray}
where $ \rho_0$ and $ \rho_b$ are the central and surface density of the star respectively. we assume $b$ as the radius of the star .

To check the dominancy of matter density and pressure at the centre of the star we observe that, they are maximum at the centre and decreases
monotonically.

Therefore, the energy density and the
pressure are well behaved in the interior of the stellar
structure. Variations of the energy-density and  pressure of the compact stars in LMXBs EXO 1745-248, KS 1731-260 and 4U 1608-52
have been shown in Fig.~1 and Fig.~2, respectively.

\begin{figure}
\includegraphics[height=1.5in, width=1.5in]{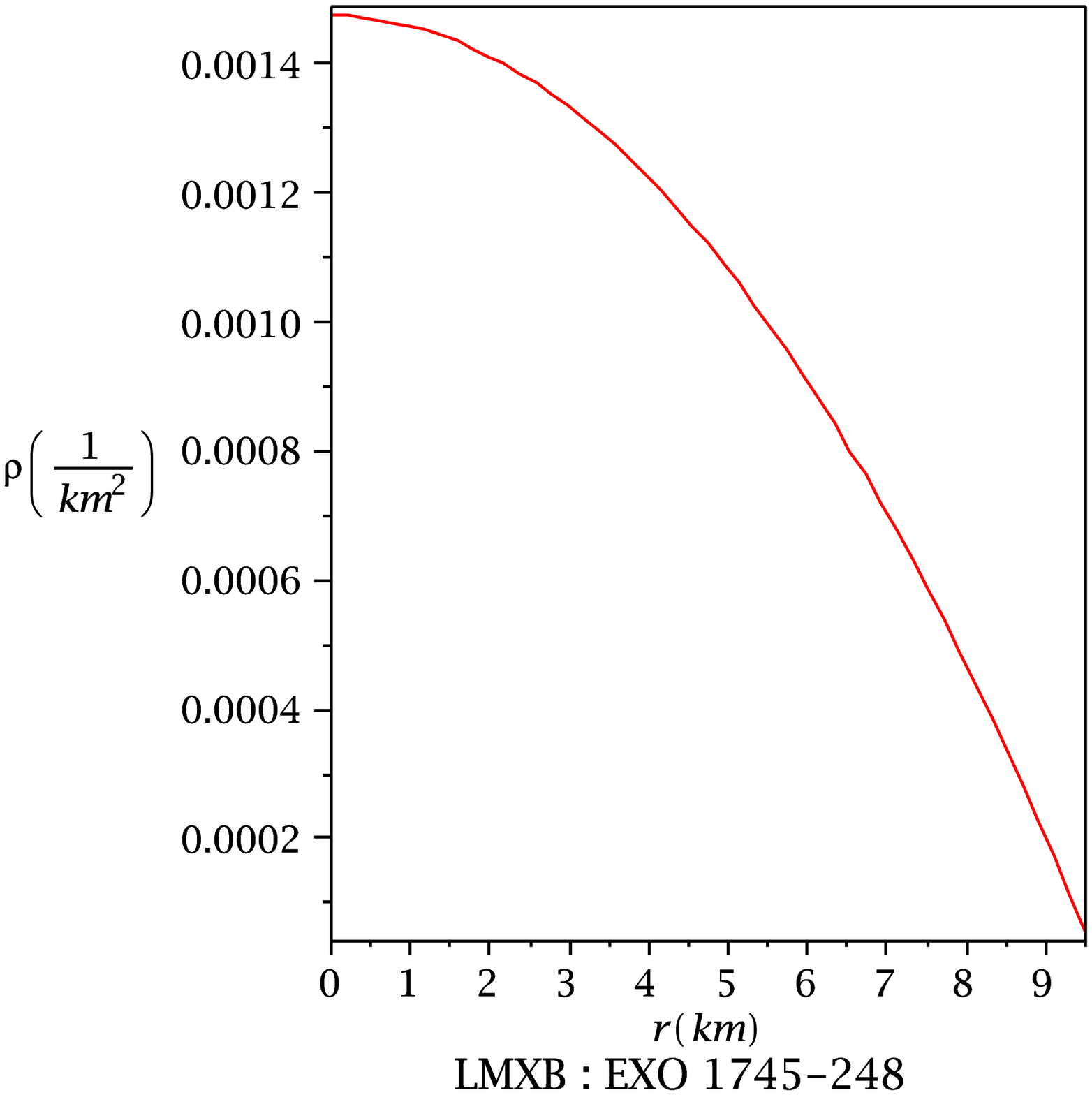}~~~~~~~
\includegraphics[height=1.5in, width=1.5in]{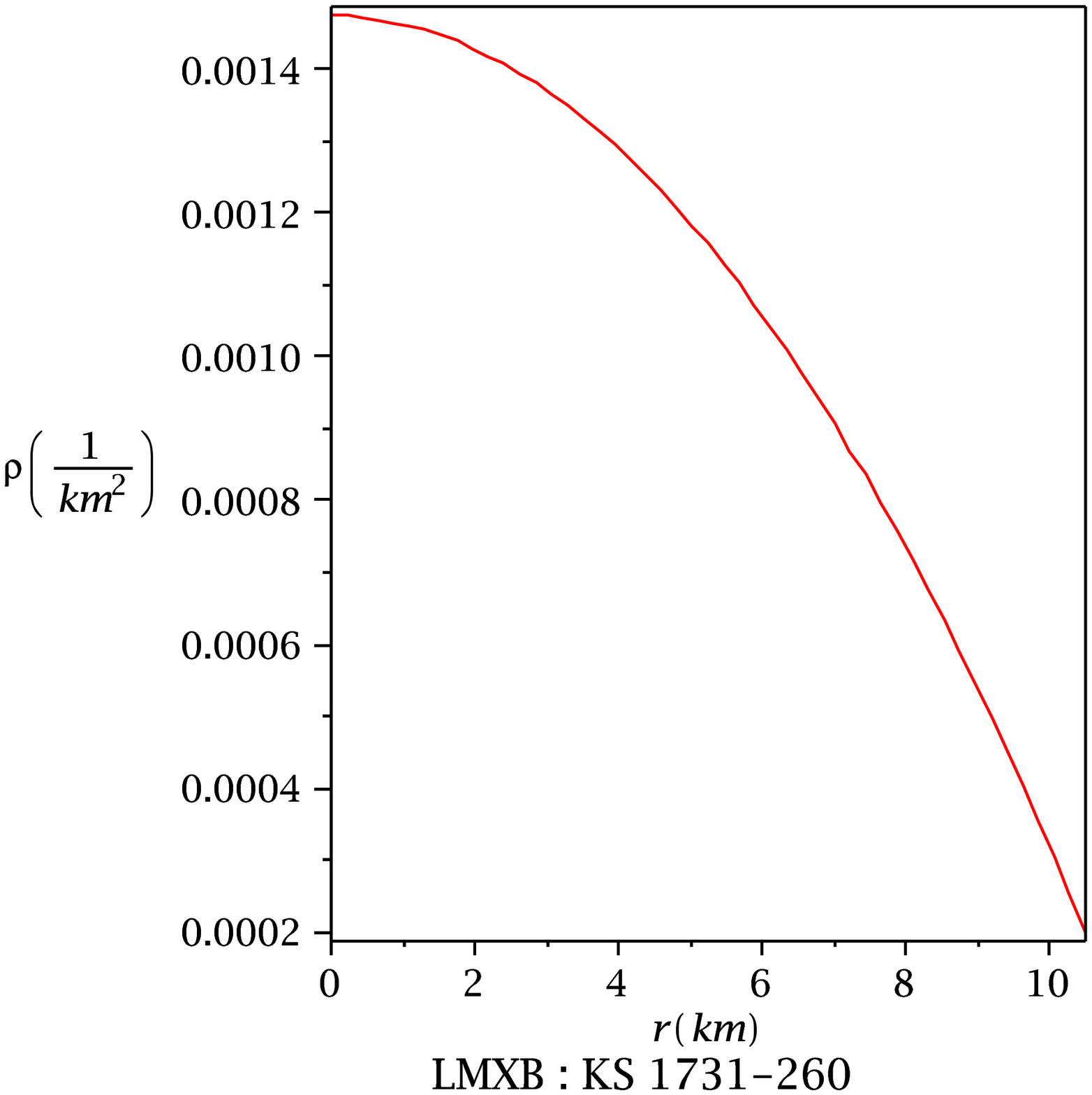}\\
\includegraphics[height=1.5in, width=1.5in]{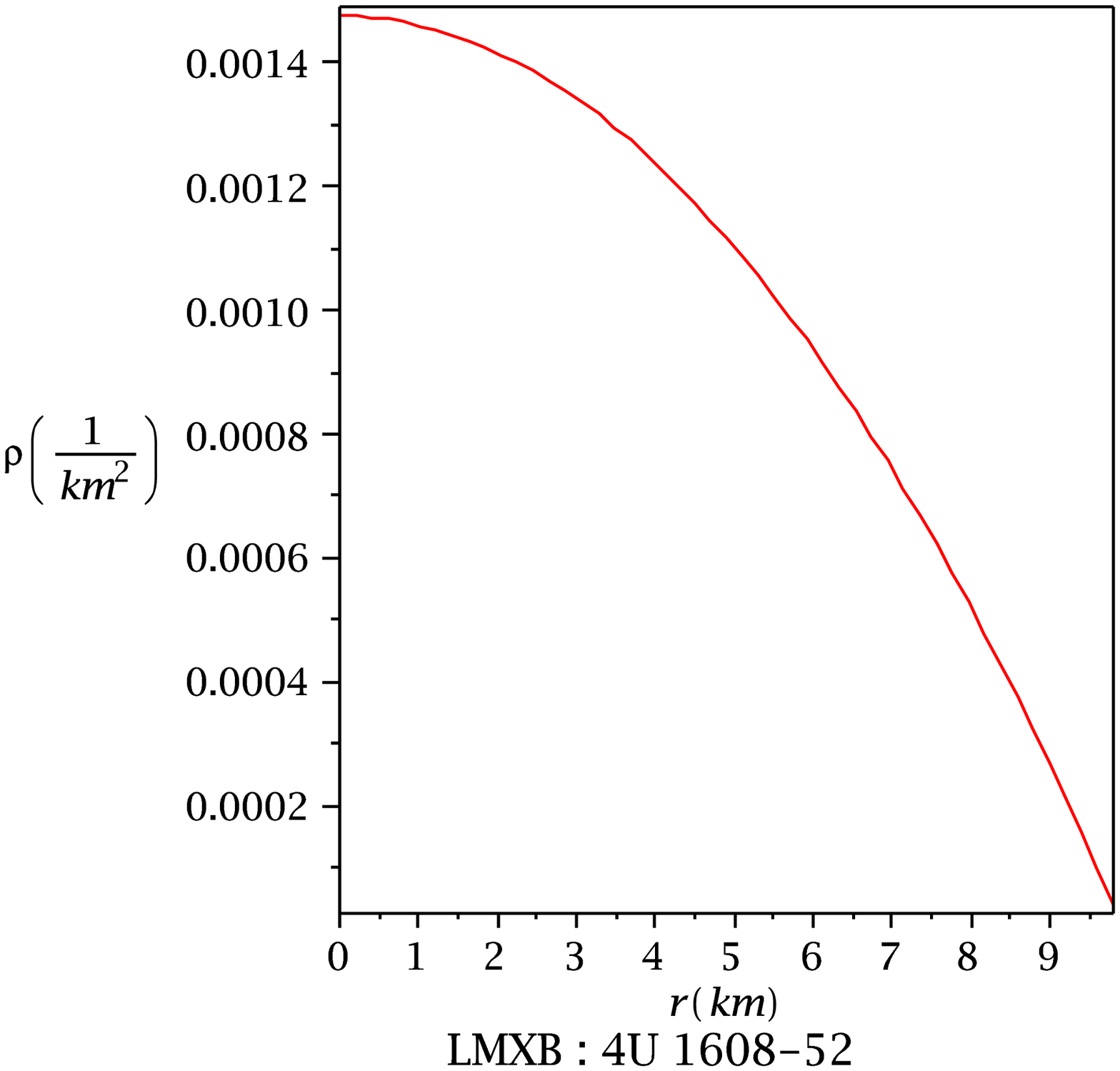}\\
FIG. 1 : Variation of the energy-density,$\rho(r)$ at the stellar interior of the compact stars in LMXBs EXO 1745-248, KS 1731-260 and 4U 1608-52 respectively. We have taken the numerical values of the parameters as R=9.0 Km,A= 15 Km, 16.2 Km and 15.2 Km respectively.
\end{figure}

\begin{figure}
\includegraphics[height=1.5in, width=1.5in]{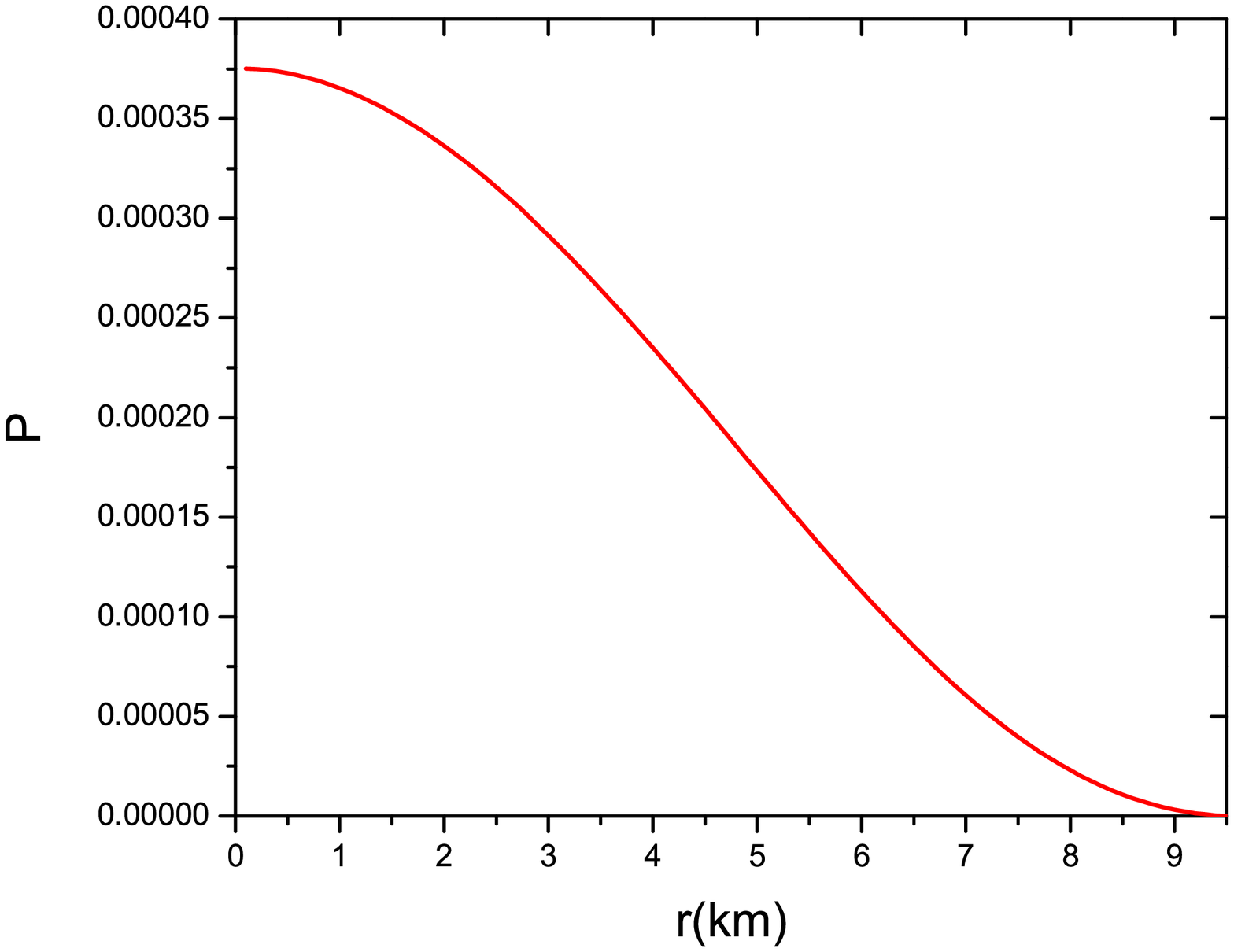}~~~~~~~
\includegraphics[height=1.5in, width=1.5in]{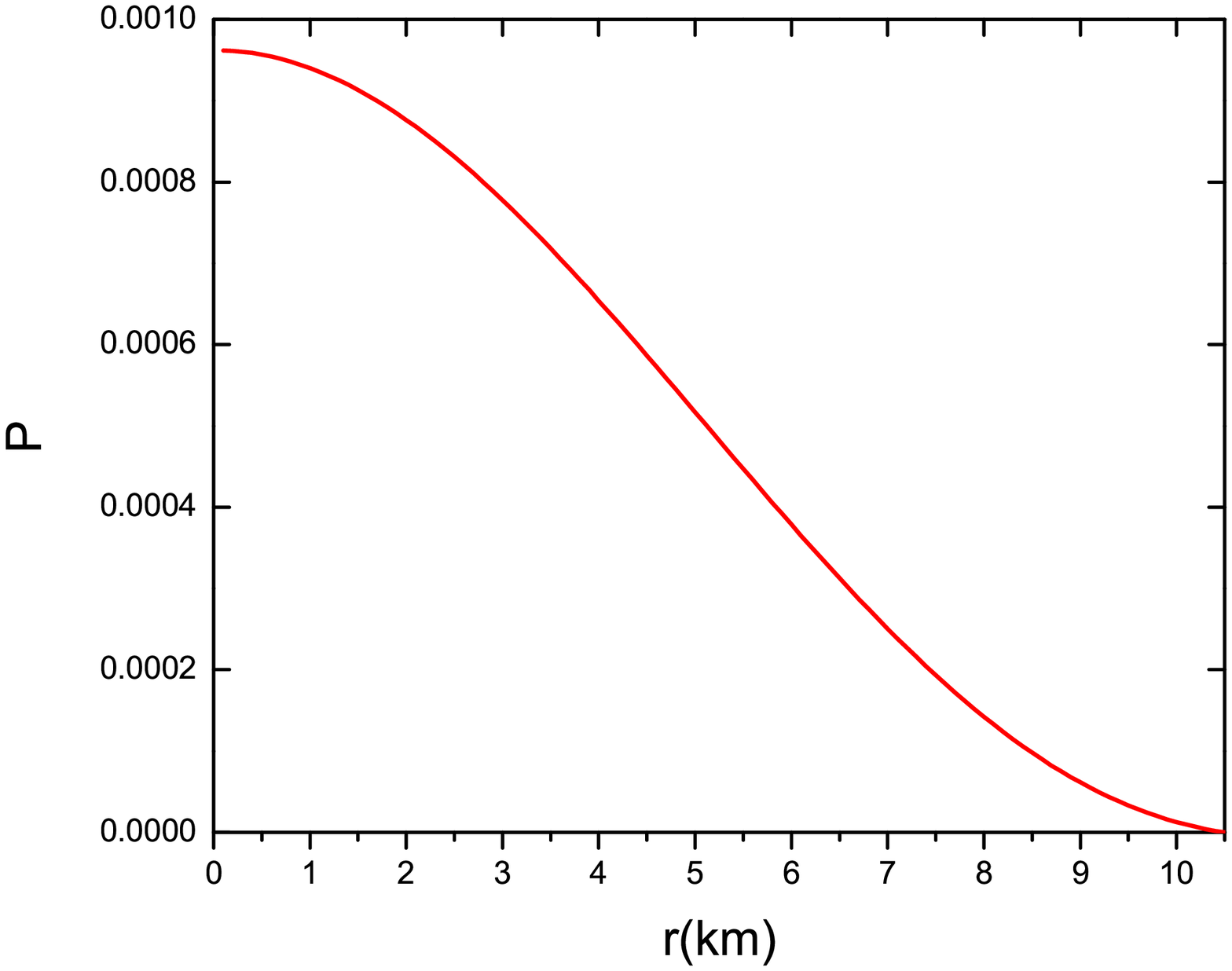}\\
\includegraphics[height=1.5in, width=1.5in]{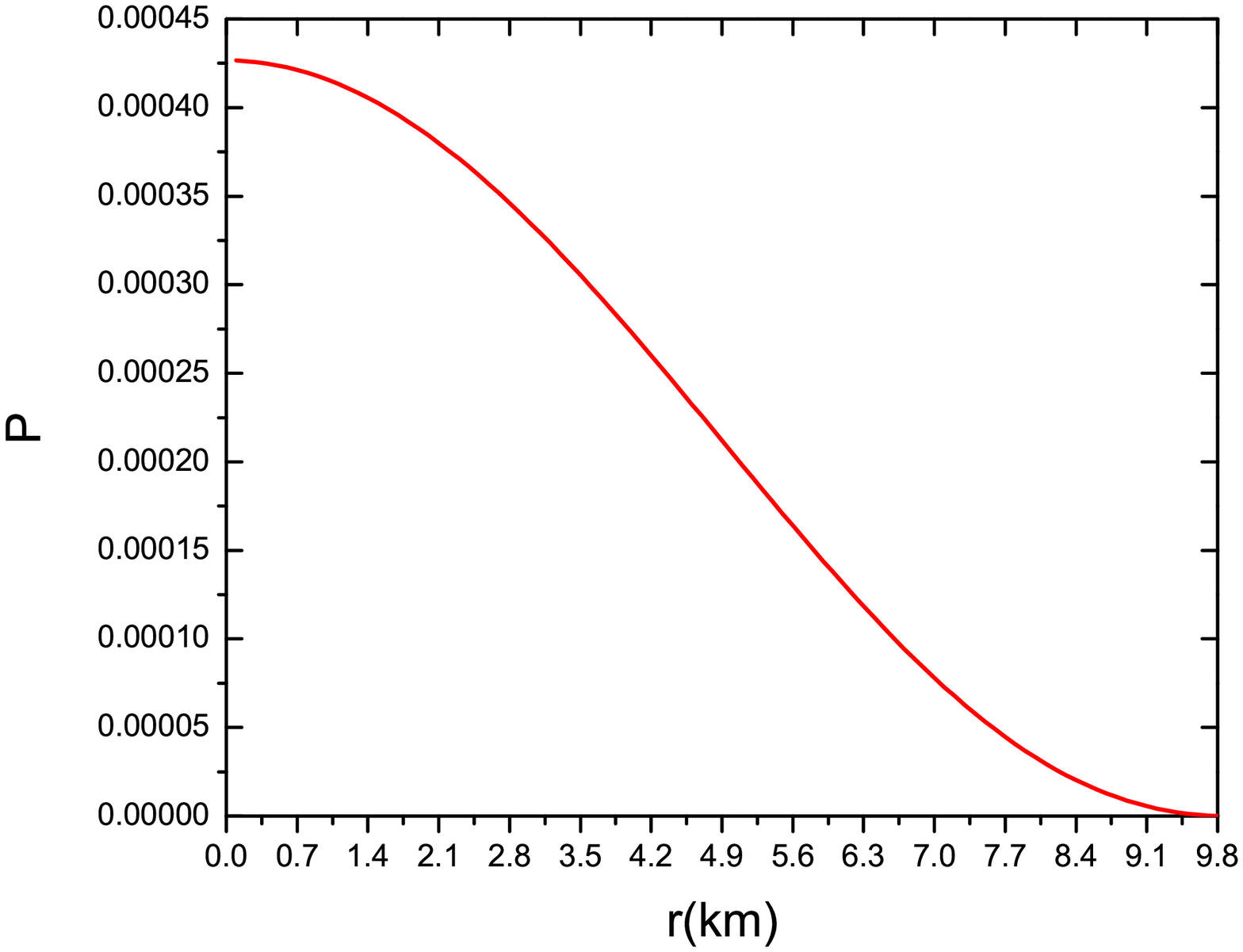}\\
FIG. 2 : Variation of the pressure ($p$) at the stellar interior of the compact stars in LMXBs EXO 1745-248, KS 1731-260 and 4U 1608-52 respectively. We have taken the numerical values of the parameters as R=9.0 Km,A= 15 Km, 16.2 Km and 15.2 Km respectively.
\end{figure}


\section{Matching Conditions}
Interior metric of the star will be matched to the Schwarzschild exterior solution
\begin{equation}
ds^2 = - \left(1-\frac{2M}{r}\right)dt^2 +  \left(1-\frac{2M}{r}\right)^{-1}dr^2 +r^2
(d\theta^2 +sin^2\theta d\phi^2), \label{eq1}
\end{equation}
at the boundary i.e., at $r=b$.
For the continuity of the metric functions $g_{tt}$, $g_{rr} $ and $\frac{\partial g_{tt}}{\partial r}$ at the boundary, we get

\begin{eqnarray}
\left(1-\frac{b^2}{R^2}+4\frac{b^4}{A^4}\right) &=& 1 - \frac{2M}{b},\label{eq13}
\end{eqnarray}

\begin{equation}
B^2\sin ^{2}\ln \sqrt{\frac{\sqrt{1-\frac{b^2}{R^2}+4\frac{b^4}{A^4}}+2\frac{b^2}{A^2}-\frac{1}{4}\frac{A^2}{R^2}}{C}} =
 \left(1-\frac{2M}{b}\right)  .\label{eq14}
\end{equation}

Now from the equation ~(\ref{eq13}) , we get the compactification factor as
\begin{equation}
u = \frac{M}{b} = \left(\frac{b^2}{2R^2}-2\frac{b^4}{A^4}\right).\label{eq15}
\end{equation}

\section{Some special features}
\subsection{TOV equation}
For fluid distribution, the generalized TOV equation has the form
\begin{equation}
\frac{dp}{dr} +\frac{1}{2} \nu^\prime\left(\rho
 + p\right)
= 0.\label{eq18}
\end{equation}

The modified TOV equation describes the equilibrium condition for the compact star subject to
effective gravitational($F_g$) and effective hydrostatic($F_h$) force nature of the stellar object as
\begin{equation}
F_h+ F_g  = 0,\label{eq21}
\end{equation}
where,
\begin{eqnarray}
F_g &=& \frac{1}{2} \nu^\prime\left(\rho+p\right)\label{eq22}\\
F_h &=& \frac{dp}{dr} \label{eq23}
\end{eqnarray}
Therefore,  the static
equilibrium configurations do exist in the presence of gravitational and hydrostatic  forces.  FIG.3 shows the equilibrium condition
under gravitational and hydrostatic forces .
\begin{figure}
\includegraphics[height=1.5in, width=1.5in]{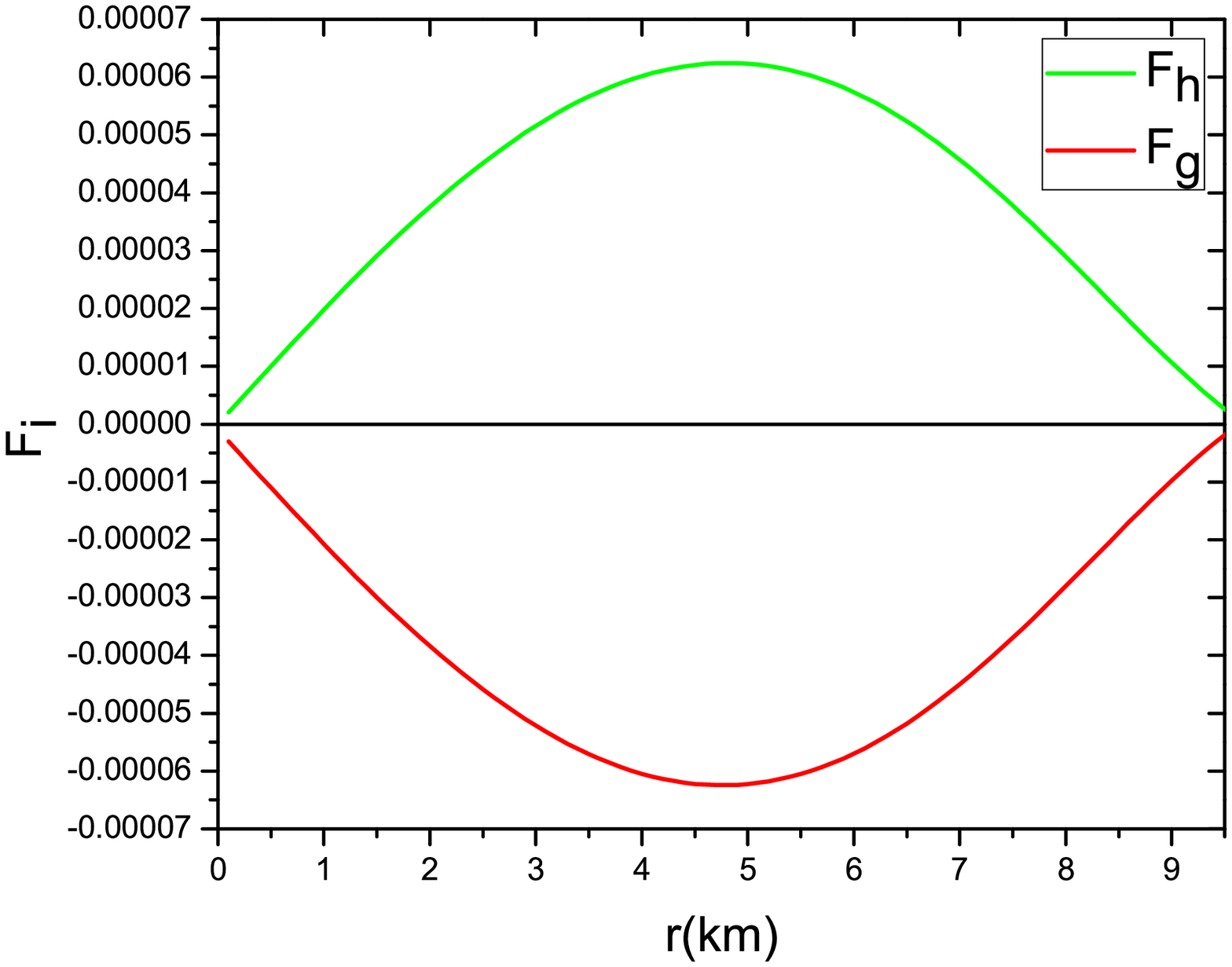}~~~~~~~
\includegraphics[height=1.5in, width=1.5in]{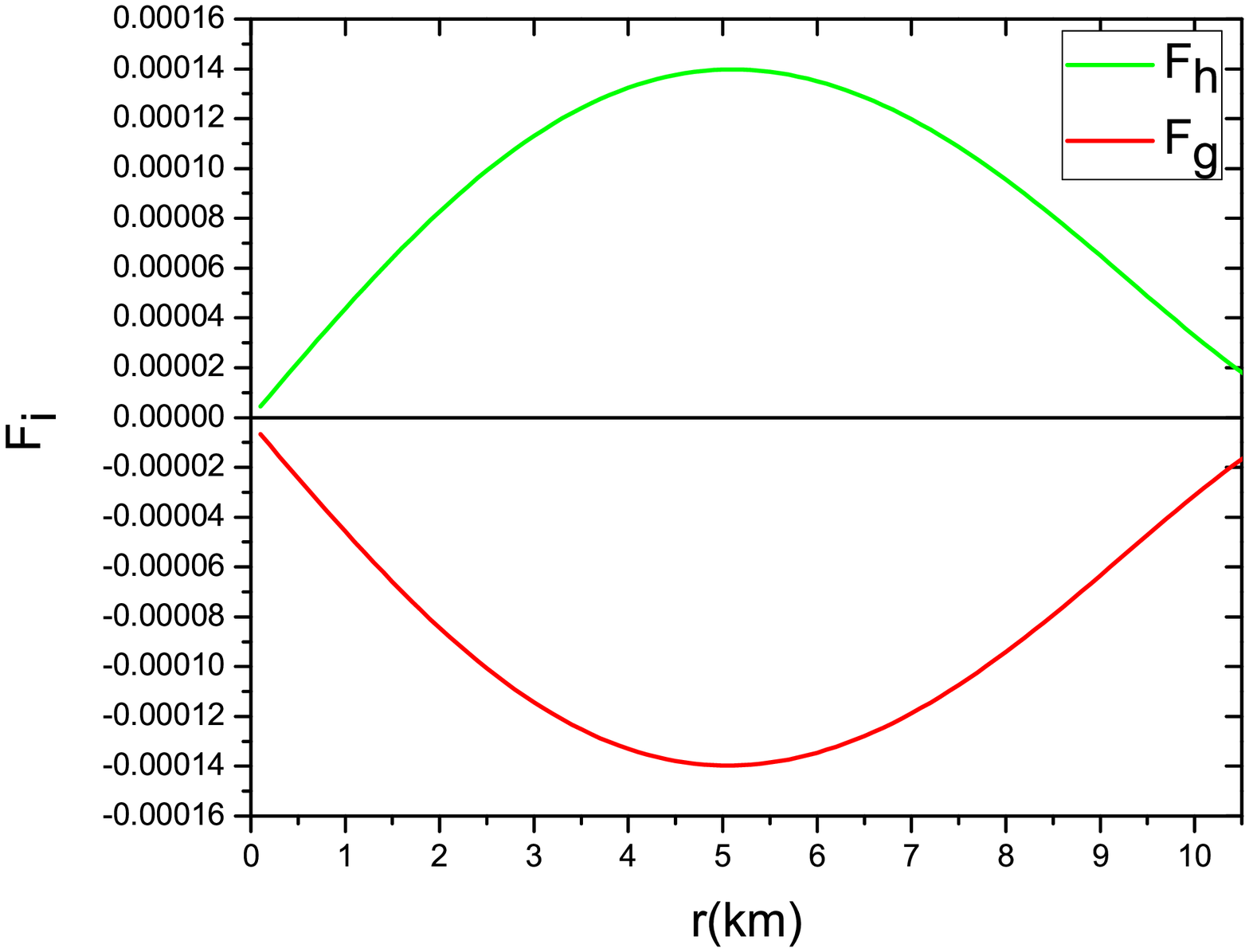}\\
\includegraphics[height=1.5in, width=1.5in]{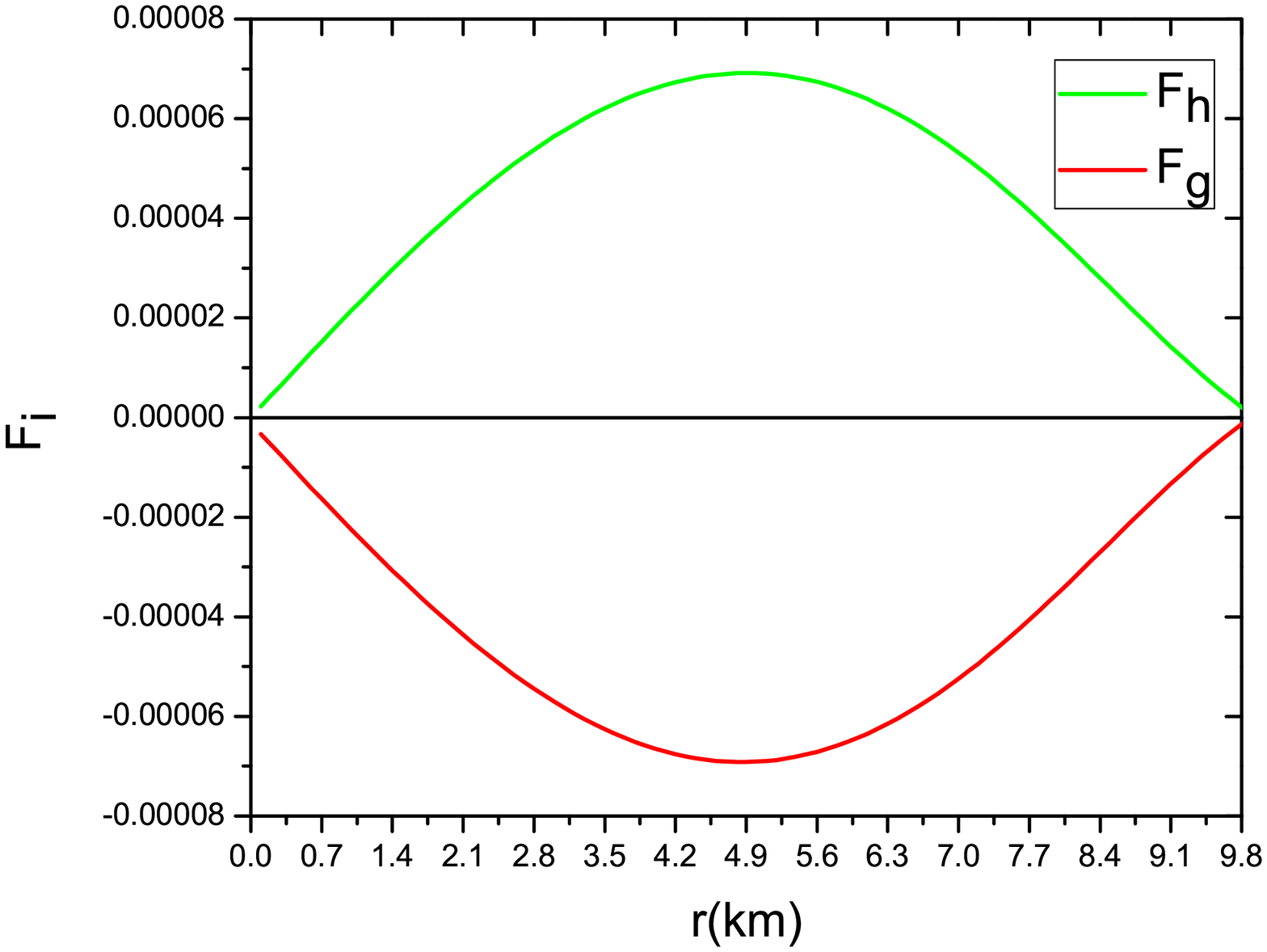}\\
FIG. 3 : Behaviors of gravitational and hydrostatic forces at the stellar interior of the compact stars in LMXBs EXO 1745-248, KS 1731-260 and 4U 1608-52 respectively. We have taken the numerical values of the parameters as R=9.0 Km,A= 15 Km, 16.2 Km and 15.2 Km respectively.
\end{figure}

\subsection{Energy conditions}
In our model,all the energy conditations, namely, null energy condition(NEC), weak energy condition(WEC), strong energy condition(SEC)
and dominant energy condition(DEC), are satisfied at the centre ($r=0$). It is evident from FIG. 1, FIG. 2 and the table given below
  the following energy conditions holds good:\\
(i) NEC: $p_{0}+\rho_{0}\geq0$ ,\\
(ii) WEC: $p_{0}+\rho_{0}\geq0$  , $~~\rho_{0}\geq0$  ,\\
(iii) SEC: $p_{0}+\rho_{0}\geq0$  ,$~~~~3p_{0}+\rho_{0}\geq0$ ,\\
(iv) DEC: $\rho_{0} > |p_{0}| $.

\begin{figure}
\includegraphics[height=1.5in, width=1.5in]{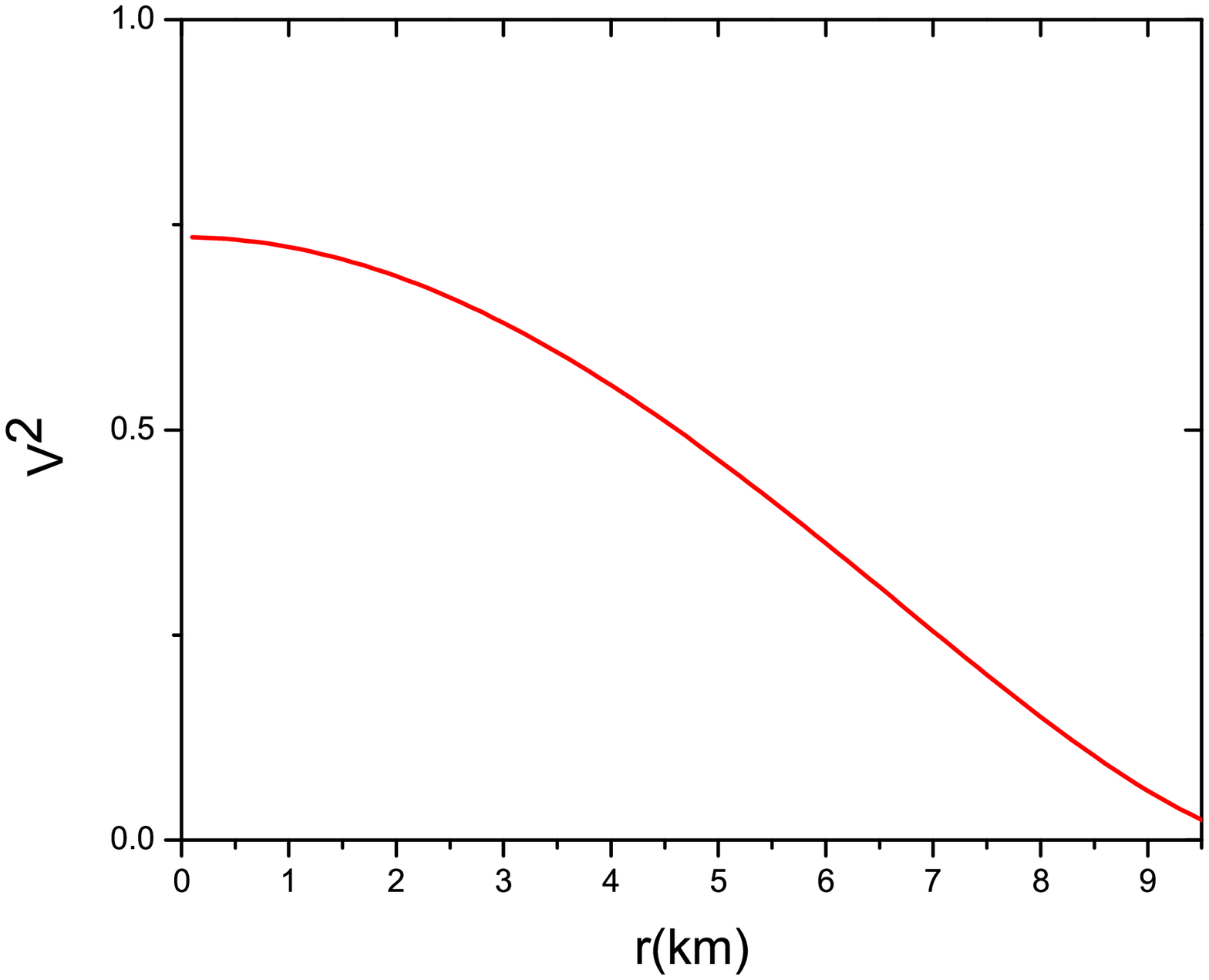}~~~~~~~
\includegraphics[height=1.5in, width=1.5in]{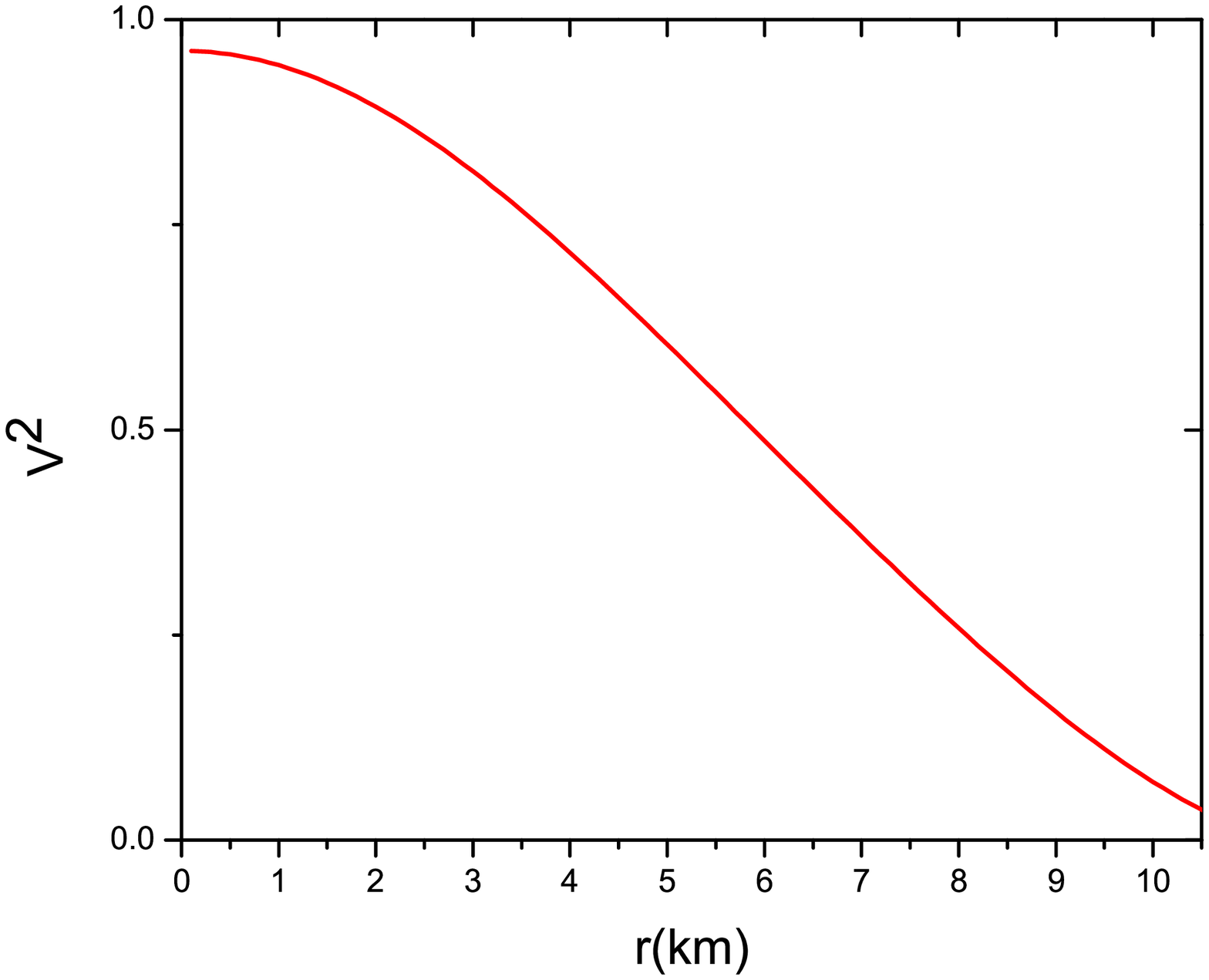}\\
\includegraphics[height=1.5in, width=1.5in]{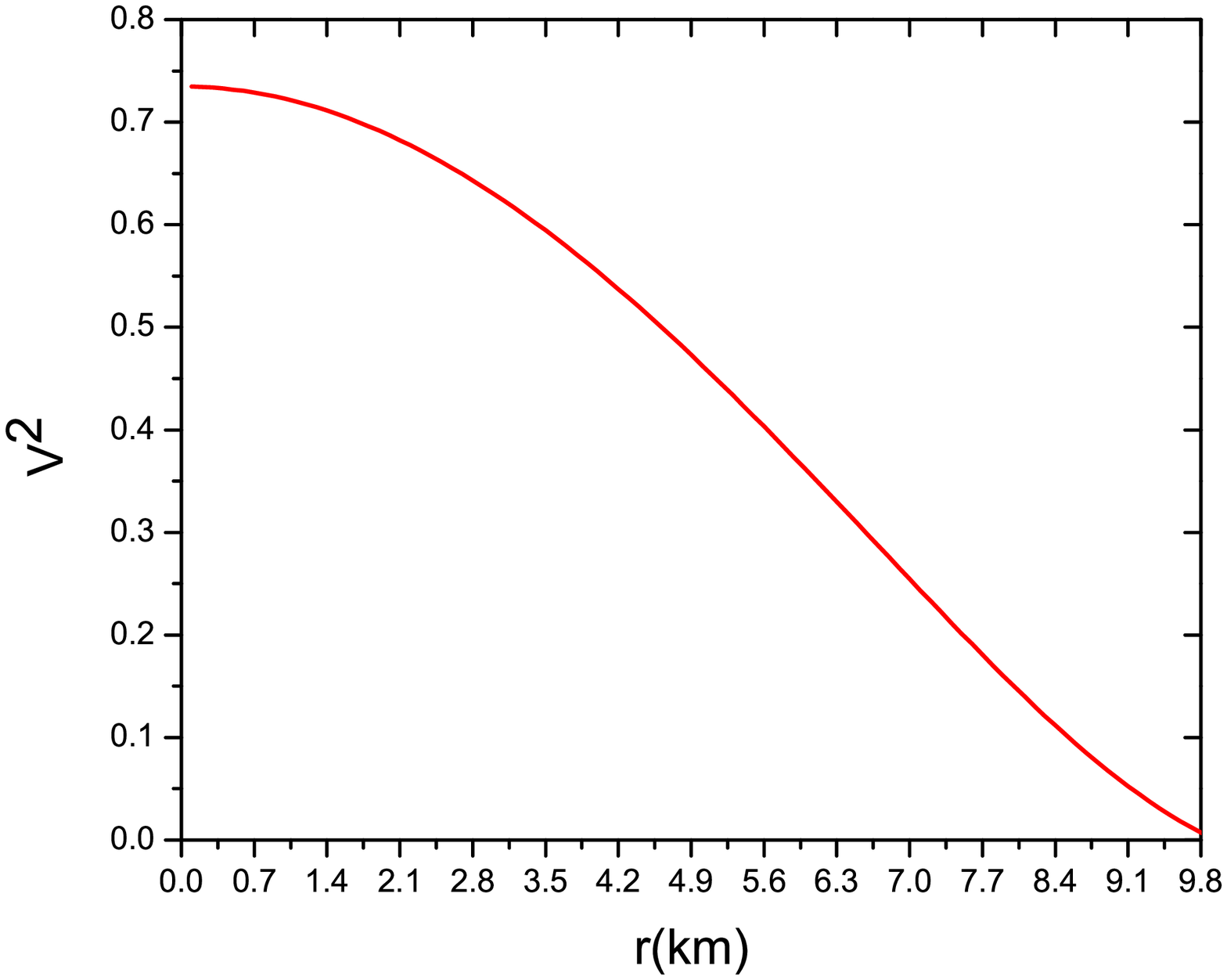}\\
FIG. 4 : Variation of sound speed of the compact stars in LMXBs EXO 1745-248, KS 1731-260 and 4U 1608-52 respectively. We have taken the numerical values of the parameters as R=9.0 Km,A= 15 Km, 16.2 Km and 15.2 Km respectively.
\end{figure}

\subsection{Stability}
For a physically acceptable model, velocity of sound should be within the range
$0 \leq  v^2=(\frac{dp}{d\rho})\leq 1$\citep{Herrera1992,Abreu2007}.
  In our isotropic model, $ v^{2} \leq 1,$. We plot the sound speed of the compact stars in LMXBs EXO 1745-248, KS 1731-260 and 4U 1608-52 respectively in FIG.4 and observed that it satisfies well the inequalities $0\leq v^2 \leq 1$. Therfore our model is well stabled.


\subsection{Mass-Radius relation and Surface redshift}
Here, we studied the maximum allowable mass-radius ratio. According to Buchdahl \citep{Buchdahl1959}, for a
static spherically symmetric perfect fluid, allowable mass-radius
ratio is given by $\frac{ Mass}{Radius} < \frac{4}{9}$.
In our model, the gravitational mass (M) in terms of the energy density
$\rho$ can be expressed as
\begin{equation}
\label{eq34}
 M=4\pi\int^{b}_{0} \rho~~ r^2 dr =
 \frac{b}{2}\left[\frac{b^2}{R^2}-4\frac{b^4}{A^4}\right]
\end{equation}

 The compactness, u is given by
\begin{equation}
\label{eq35} u= \frac{ M(b)} {b}=
 \frac{1}{2}\left[\frac{b^2}{R^2}-4\frac{b^4}{A^4}\right]
\end{equation}
The nature of the Mass and Compactness of the star are shown in FIG. 5 and FIG.6. We also measured the mass of the star in different LMXBs (shown in the table), which are very much consistent with the reported data \citep{Guver2010,Ozel2012a,Ozel2009b}.

\begin{figure}
\includegraphics[height=1.5in, width=1.5in]{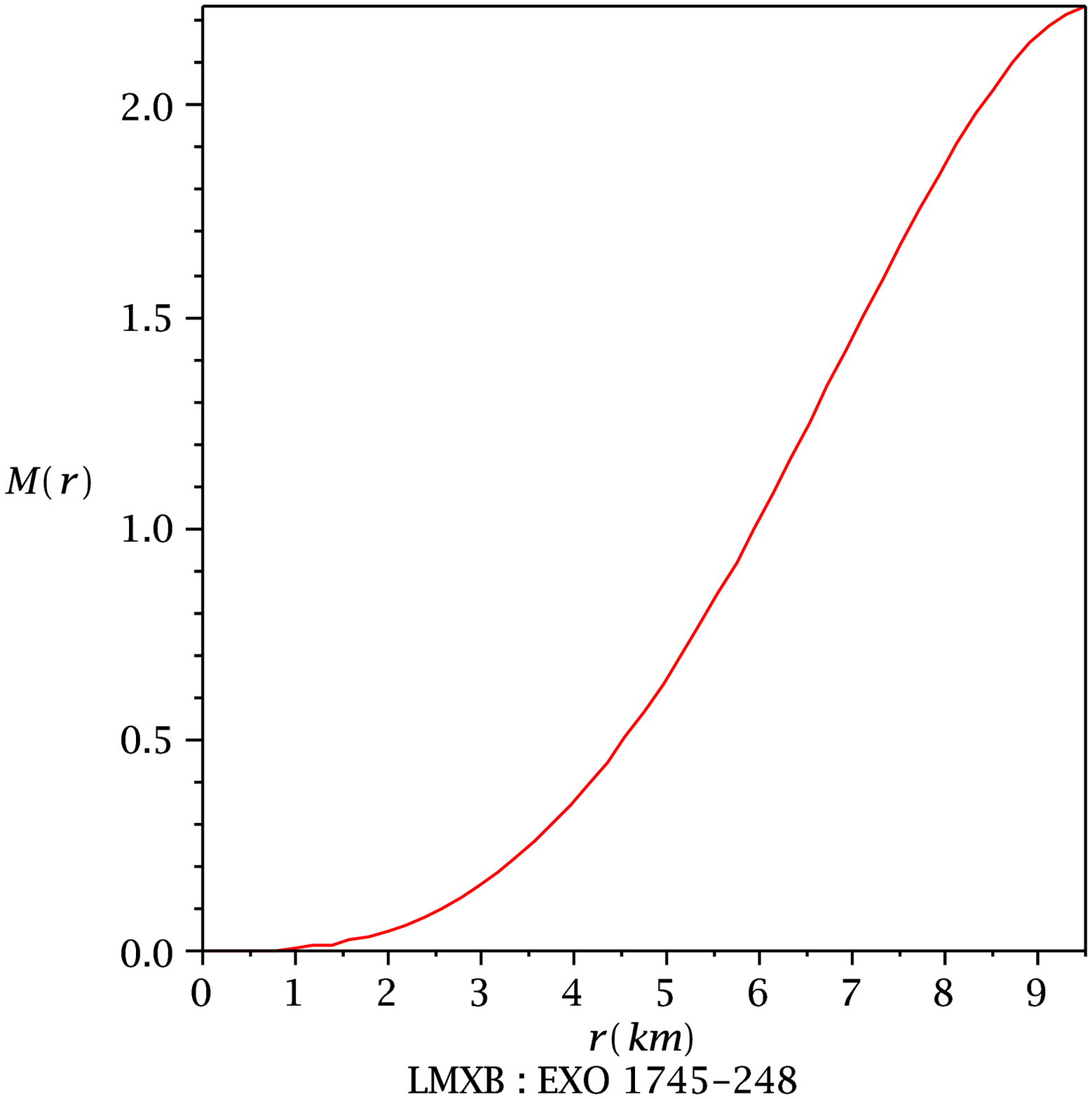}~~~~~~~
\includegraphics[height=1.5in, width=1.5in]{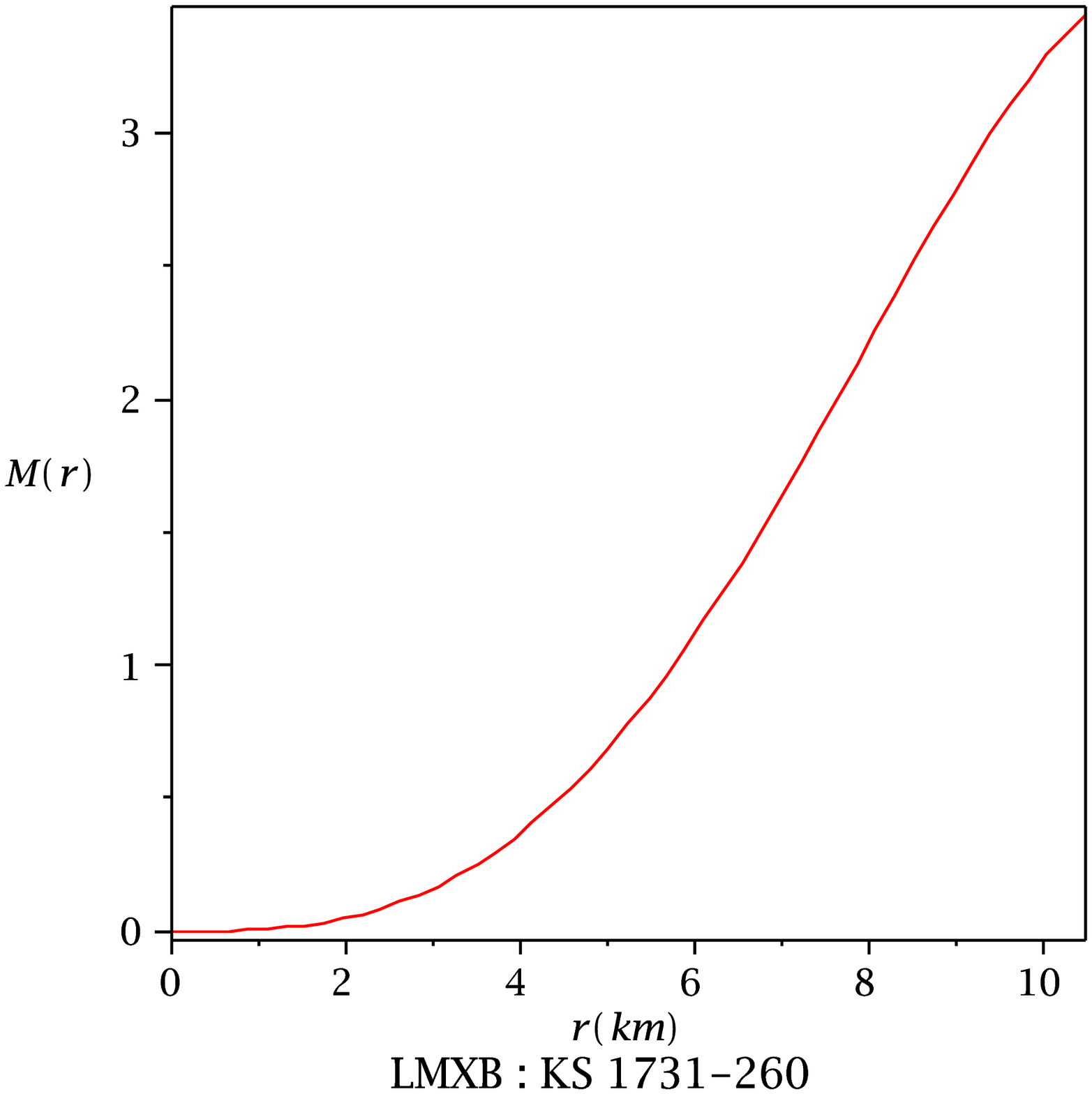}\\
\includegraphics[height=1.5in, width=1.5in]{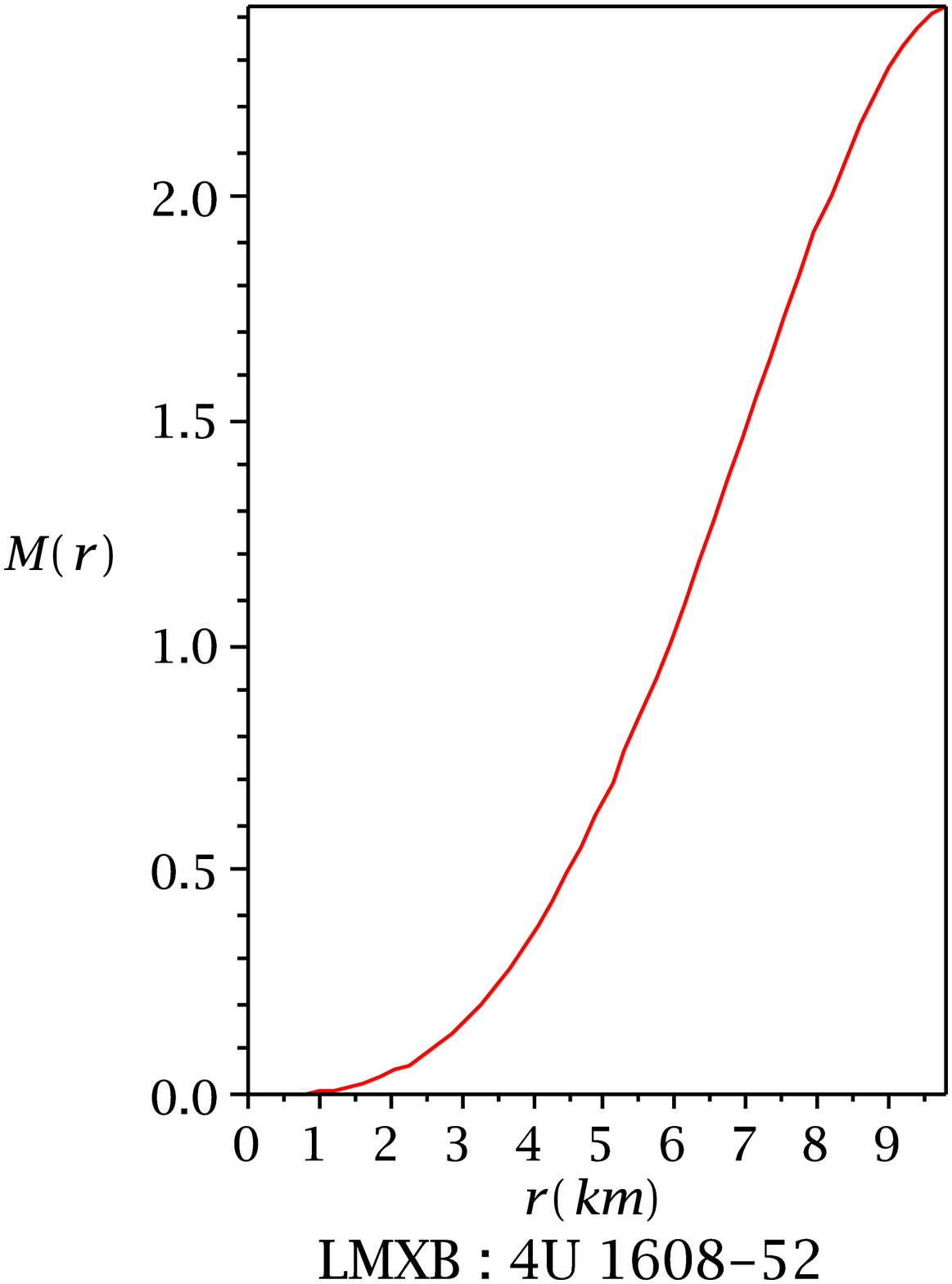}\\
FIG. 5 : Variation of Mass, M(r) at the stellar interior of the compact stars in LMXBs EXO 1745-248, KS 1731-260 and 4U 1608-52 respectively. We have taken the numerical values of the parameters as R=9.0 Km,A= 15 Km, 16.2 Km and 15.2 Km respectively.
\end{figure}
\begin{figure}
\includegraphics[height=1.5in, width=1.5in]{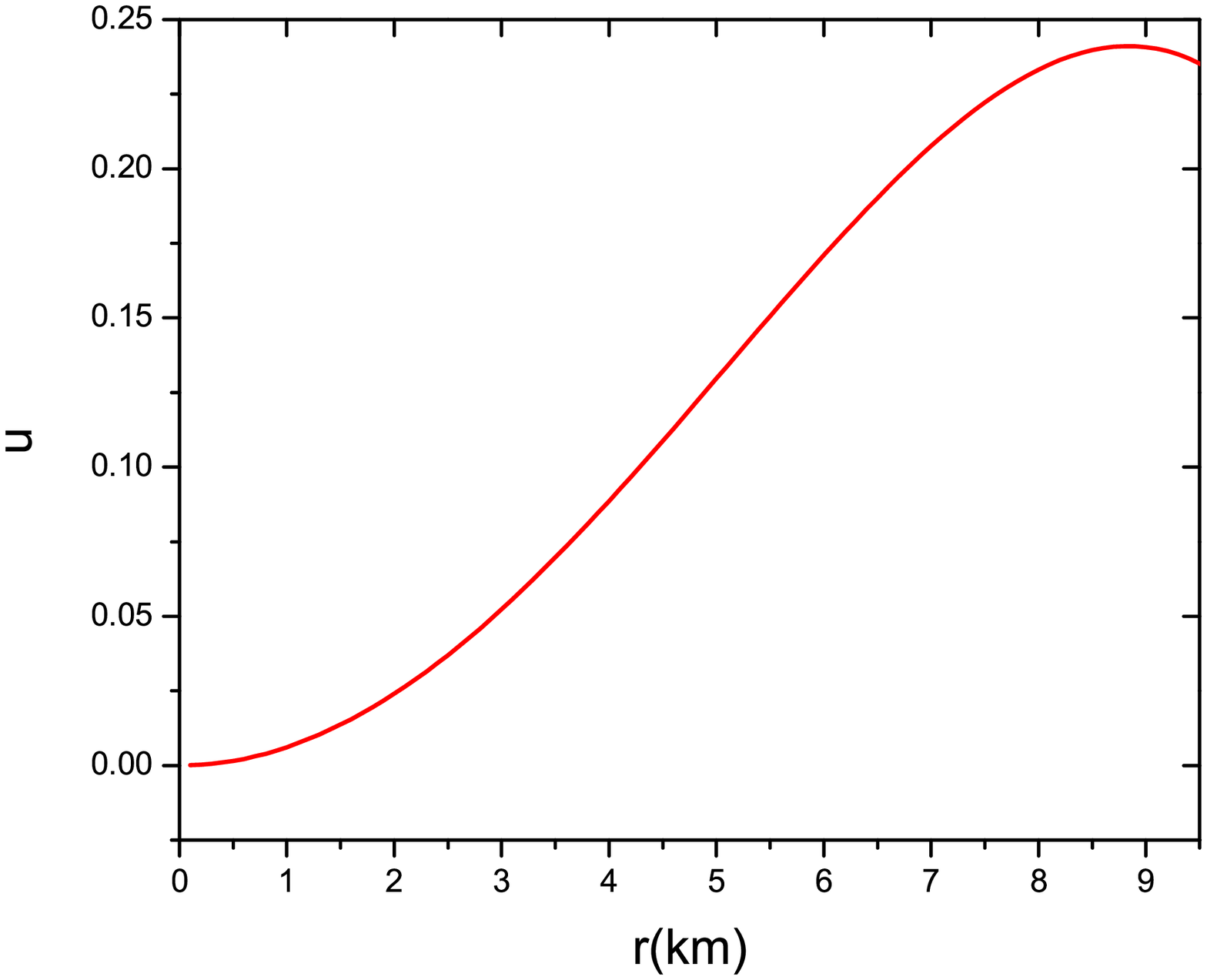}~~~~~~~
\includegraphics[height=1.5in, width=1.5in]{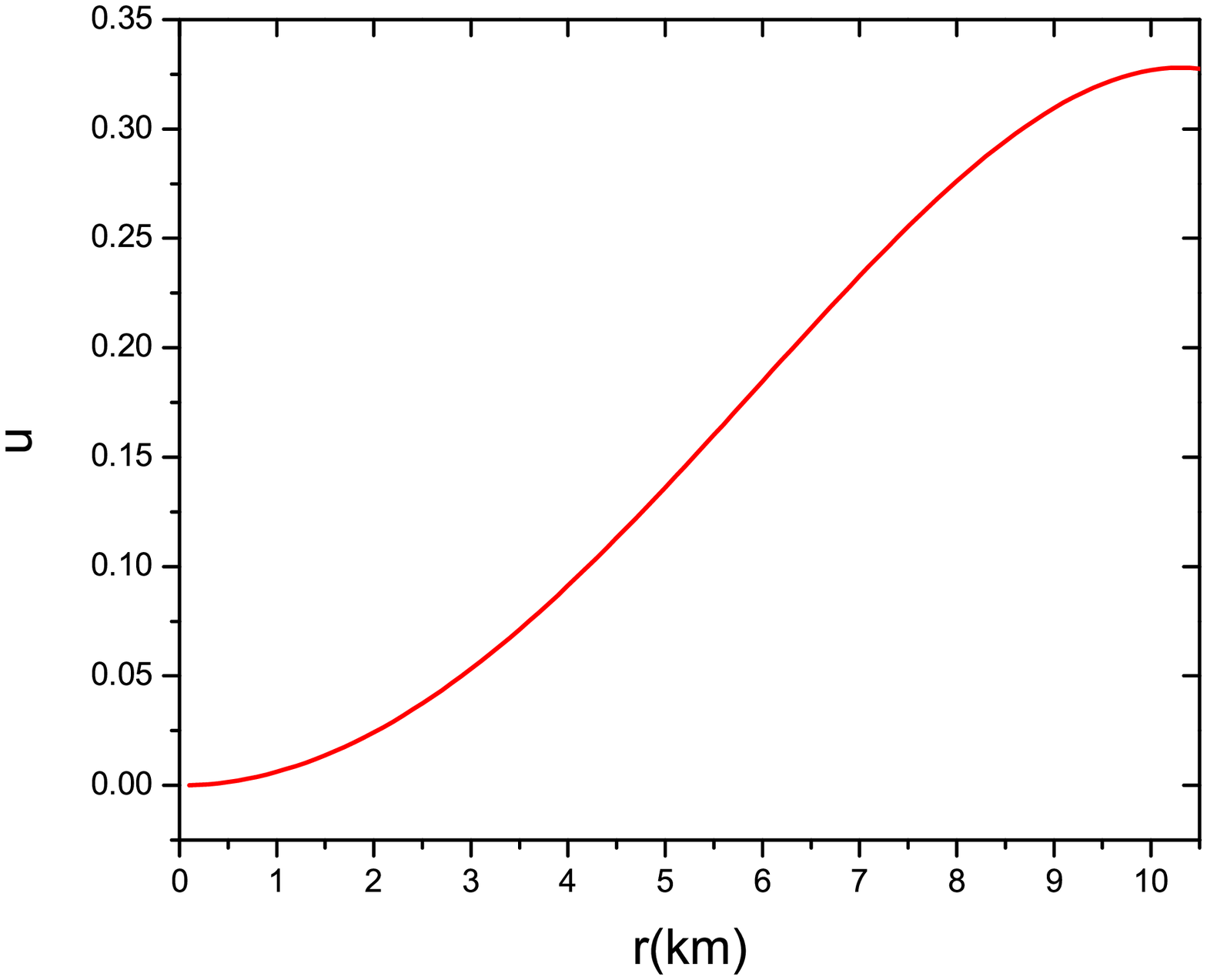}\\
\includegraphics[height=1.5in, width=1.5in]{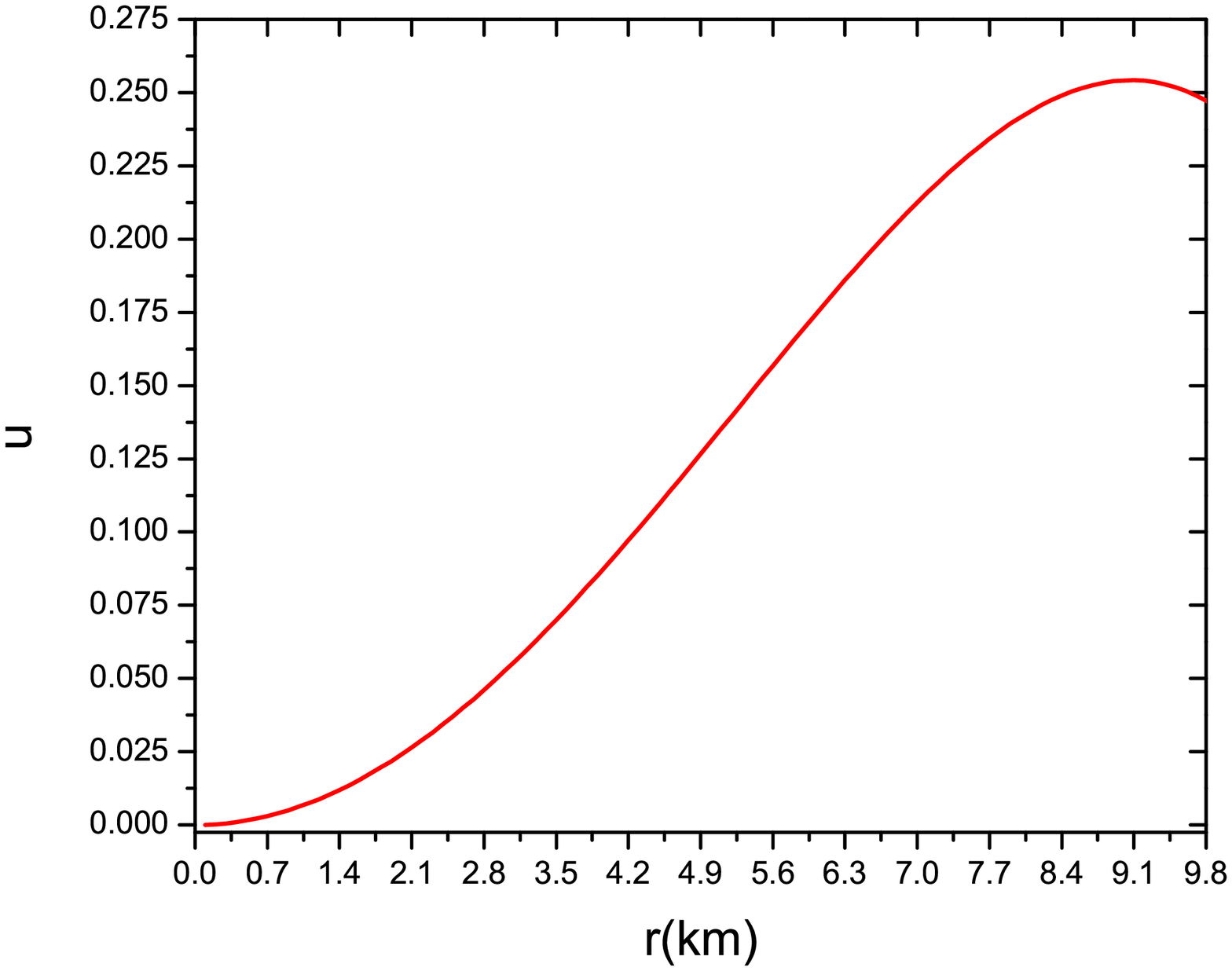}\\
FIG. 6 : Variation of Compactness of compact stars in LMXBs EXO 1745-248, KS 1731-260 and 4U 1608-52 respectively. We have taken the numerical values of the parameters as R=9.0 Km,A= 15 Km, 16.2 Km and 15.2 Km respectively.
\end{figure}


\begin{figure}
\includegraphics[height=1.5in, width=1.5in]{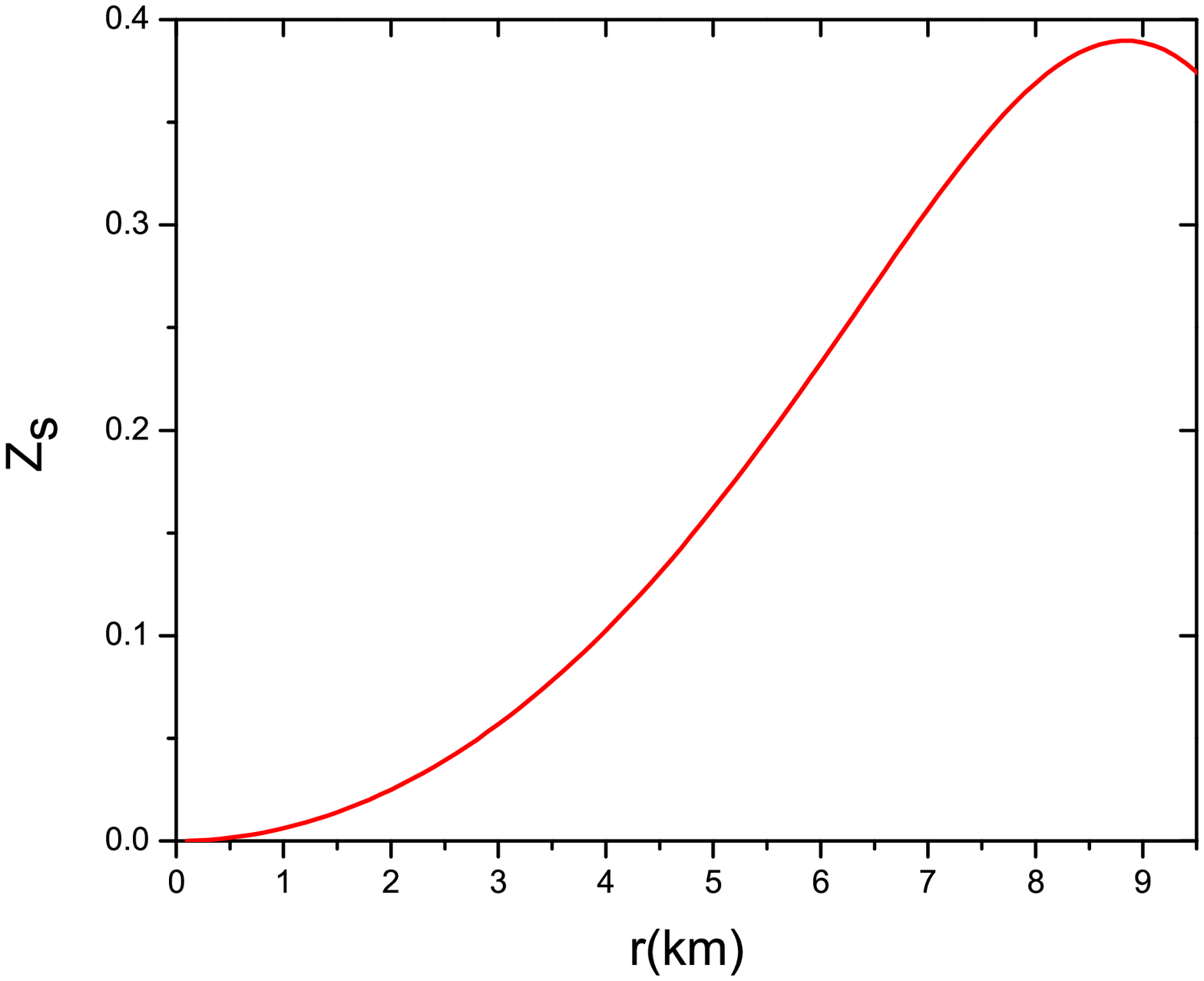}~~~~~~~
\includegraphics[height=1.5in, width=1.5in]{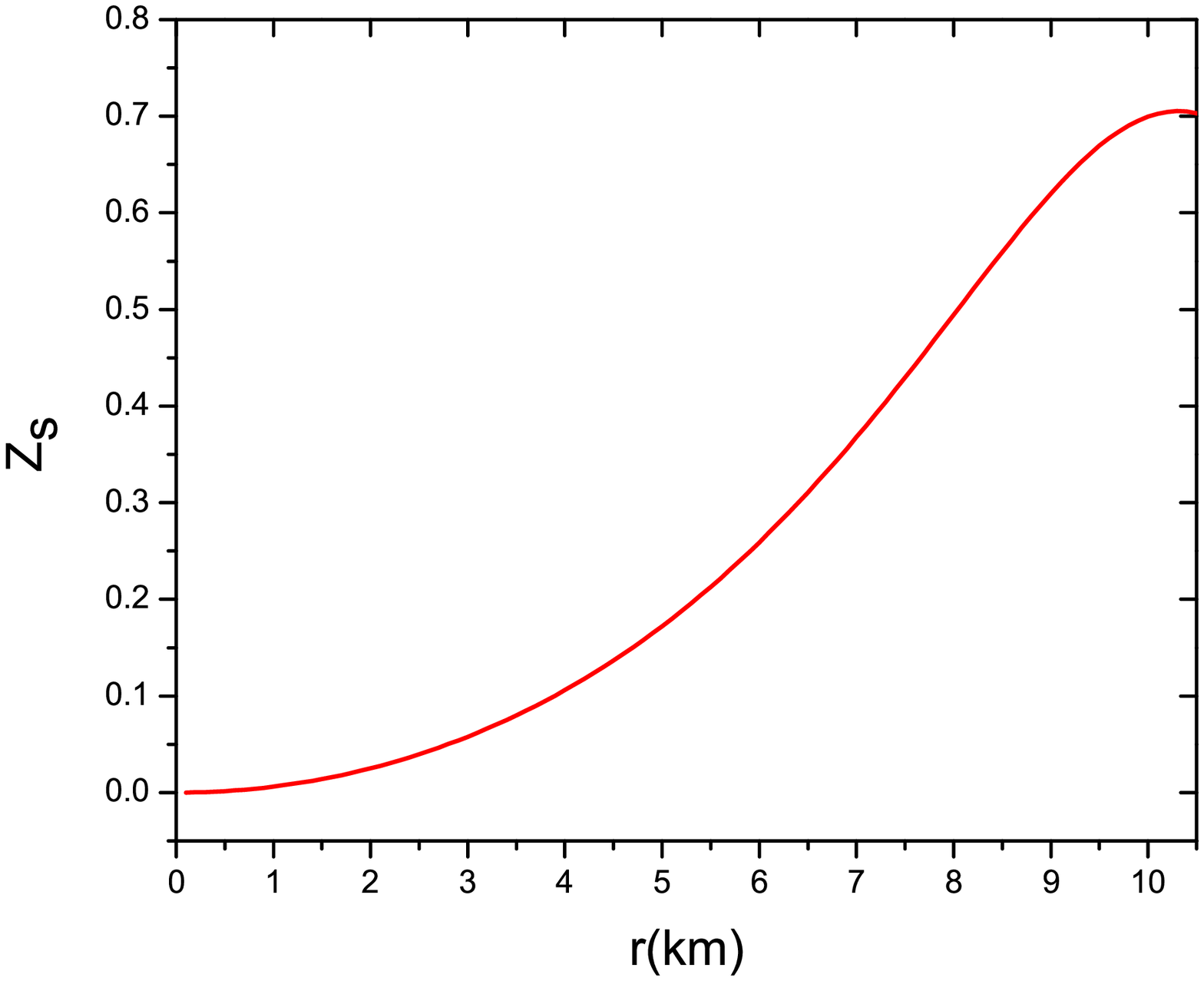}\\
\includegraphics[height=1.5in, width=1.5in]{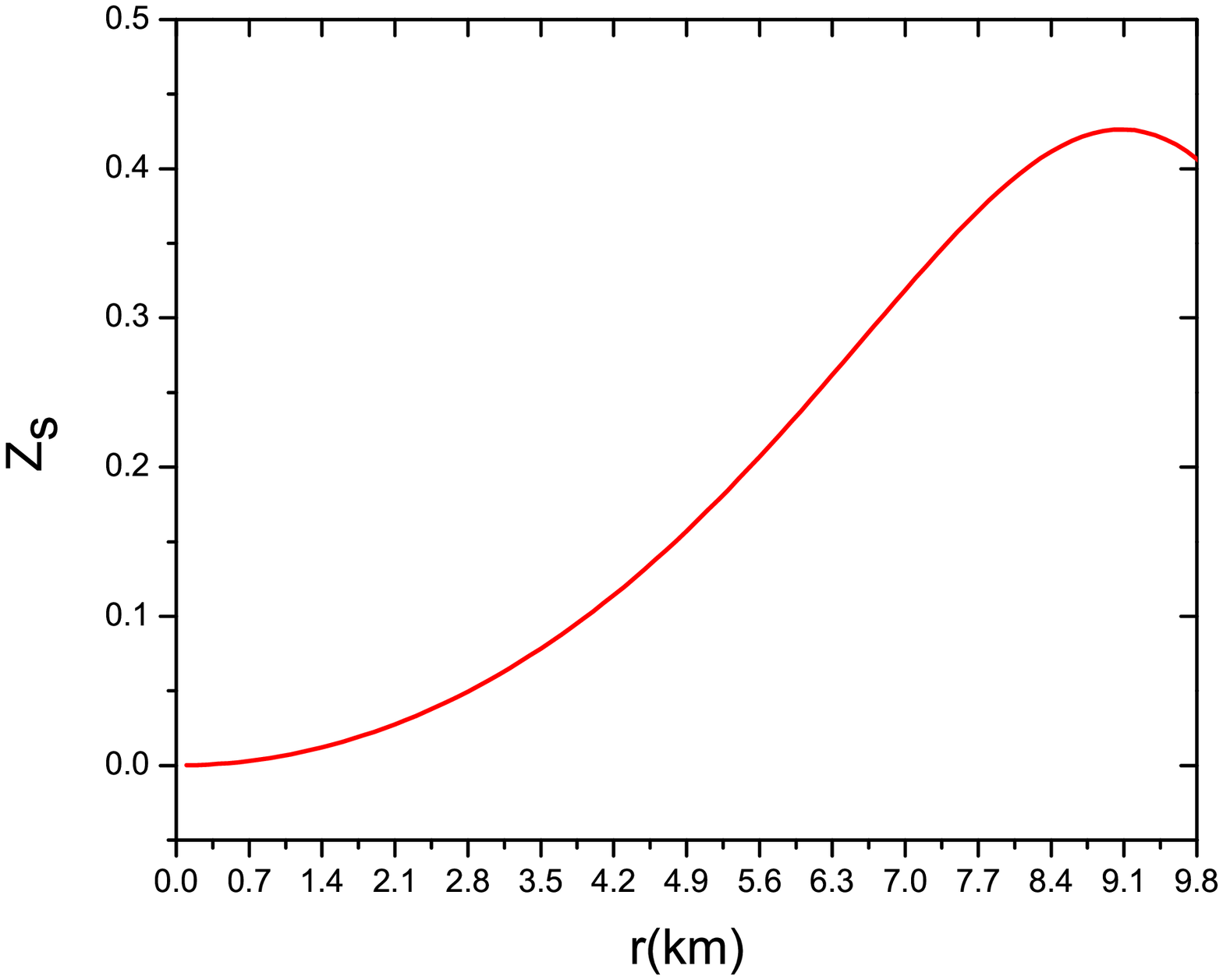}\\
FIG. 7 : Variation of red-shift function  of the compact stars in LMXBs EXO 1745-248, KS 1731-260 and 4U 1608-52 respectively. We have taken the numerical values of the parameters as R=9.0 Km,A= 15 Km, 16.2 Km and 15.2 Km respectively.
\end{figure}
\begin{table*}
\centering
\begin{minipage}{140mm}
\caption{Values of the model parameters for
different Compact Stars in LMXBs .}\label{tbl}
\begin{tabular}{@{}lrrrrrrr@{}}
\hline
Compact Star in LXMB &  $R$(km) & $Radius,b$(km) & $A$ (km) & $Mass$($M_{\odot}$) & $\rho_{0}$(gm/cc) & $\rho_{b}$(gm/cc) &$Z_s$ \\ \hline
EXO 1745-248           & 9.0       & 9.5         & 15           & 1.5234        & 1.989 x 10$^{15}$      & 0.07345 x 10$^{15}$      & 0.3744\\
KS 1731-260            & 9.0       & 10.5        & 16.2         & 2.34          & 1.989 x 10$^{15}$      & 0.26678 x 10$^{15}$      &0.7029 \\
4U 1608-52             & 9.0       & 9.8         & 15.2         & 1.65          & 1.989 x 10$^{15}$      & 0.05594 x 10$^{15}$      &0.4065\\
\hline
\end{tabular}
\end{minipage}
\end{table*}
The surface redshift ($Z_s$) corresponding to the above
compactness ($u$) is as follows:
\begin{equation}
\label{eq36} 1+Z_s= \left[ 1-(2 u )\right]^{-\frac{1}{2}} ,
\end{equation}
where
\begin{equation}
\label{eq37} Z_s= \frac{1}{\sqrt{1-\frac{b^2}{R^2}+4\frac{b^4}{A^4}}}-1
\end{equation}
Therefore, the maximum surface redshift for the isotropic compact stars  of
different radius in LMXBs can be calculated very easily(see the table and  FIG. 7).

\subsection{Equation of State}
Here we will find a relation between density and pressure of the star to get an EOS of the compact star. From eqn(3) and eqn.(4) we have,
\begin{eqnarray}
 8\pi  p &=& \frac{(\frac{\nu^\prime}{r}+\frac{1}{r^2})}{(\frac{\lambda^\prime}{r}-\frac{1}{r^2})}(8\pi \rho -\frac{1}{r^2}) -\frac{1}{r^2}.
\end{eqnarray}
For each fixed r, A and R we have,
\begin{eqnarray}
 p &=& D\rho -E(D+1),
\end{eqnarray}
where $D = \frac{(\frac{\nu^\prime}{r}+\frac{1}{r^2})}{(\frac{\lambda^\prime}{r}-\frac{1}{r^2})}$ and $E = \frac{1}{8 \pi r^2} $.

\begin{figure}
\includegraphics[height=1.5in, width=1.5in]{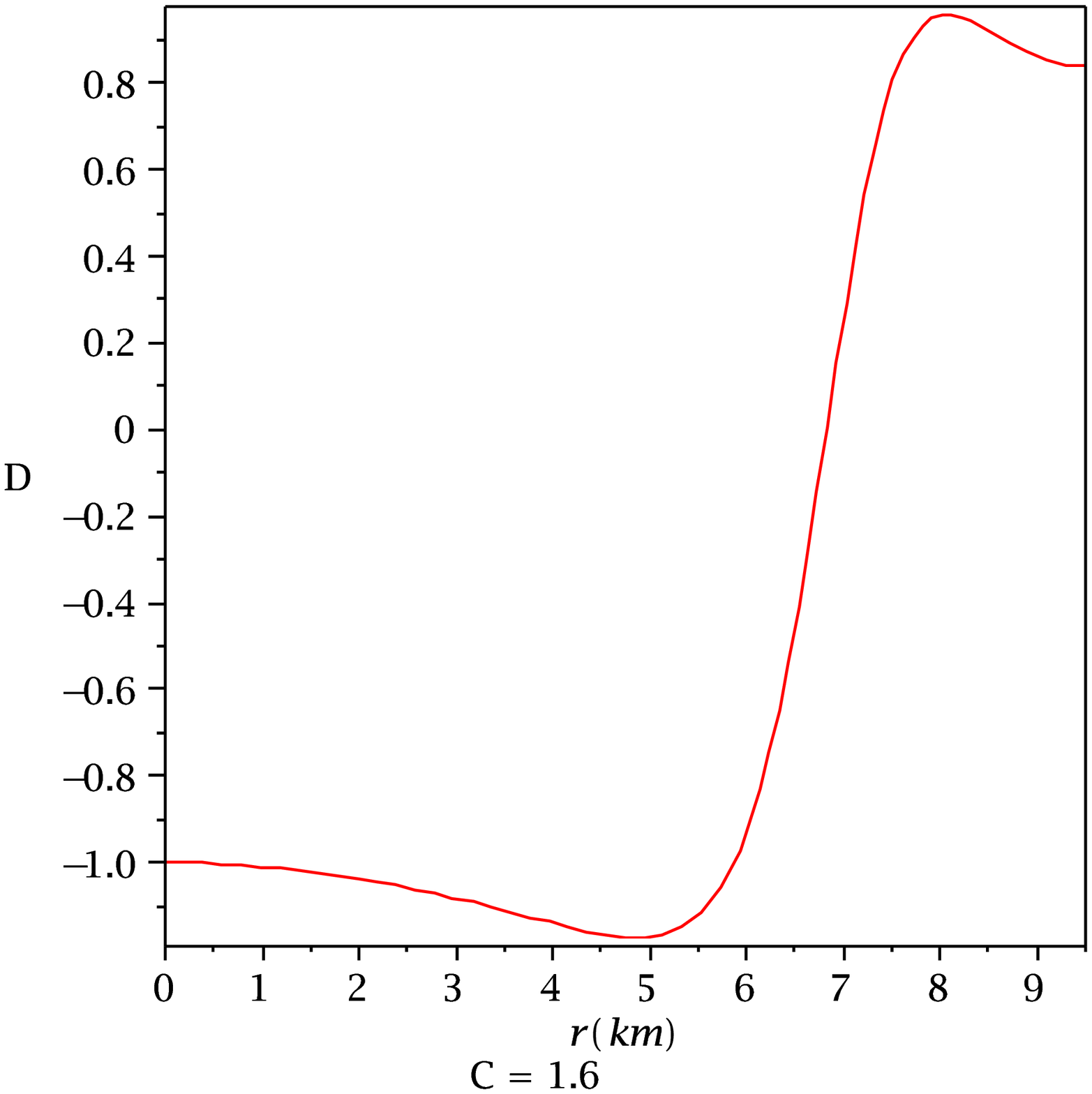}~~~~~~~
\includegraphics[height=1.5in, width=1.5in]{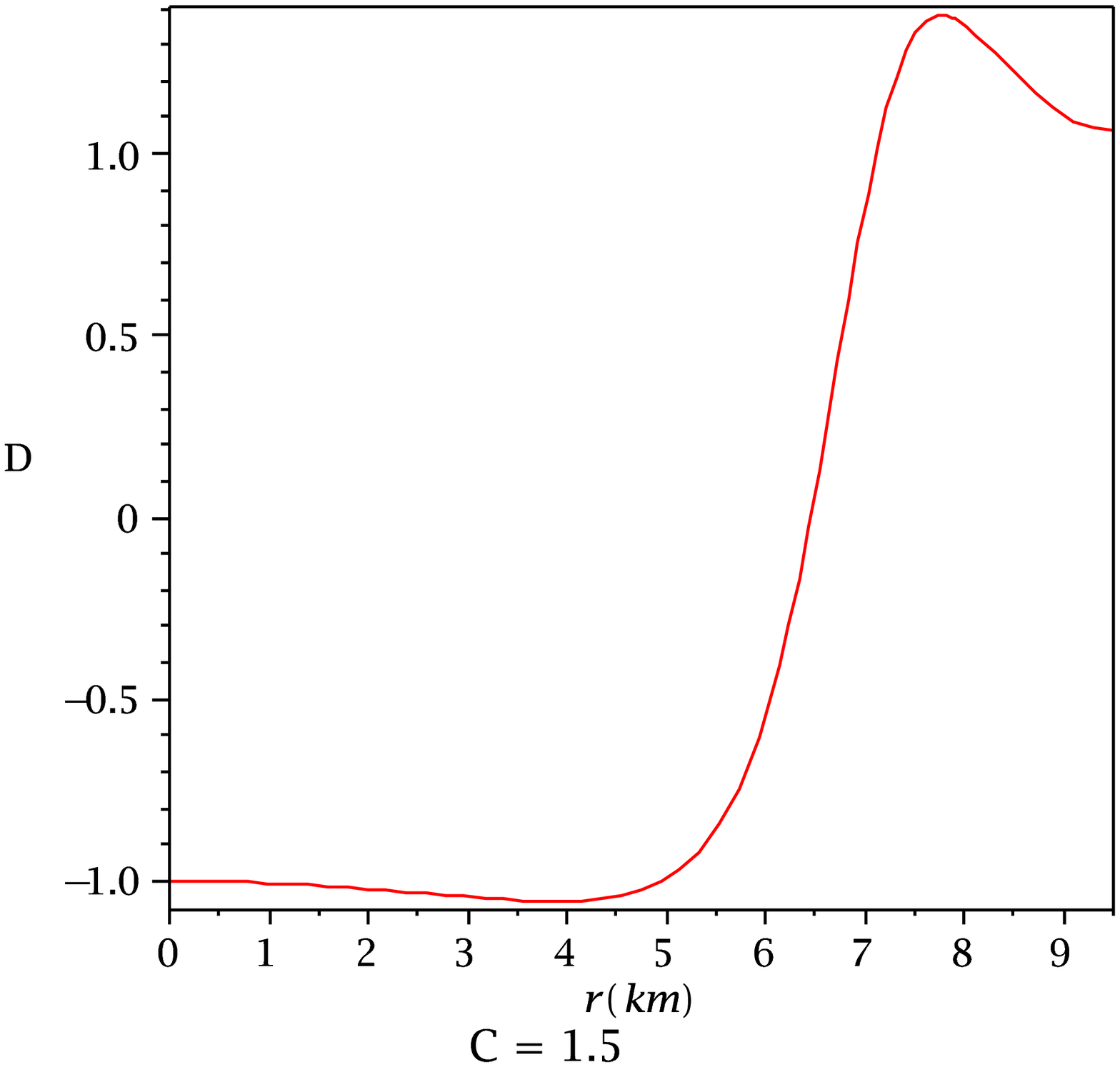}\\
\includegraphics[height=1.5in, width=1.5in]{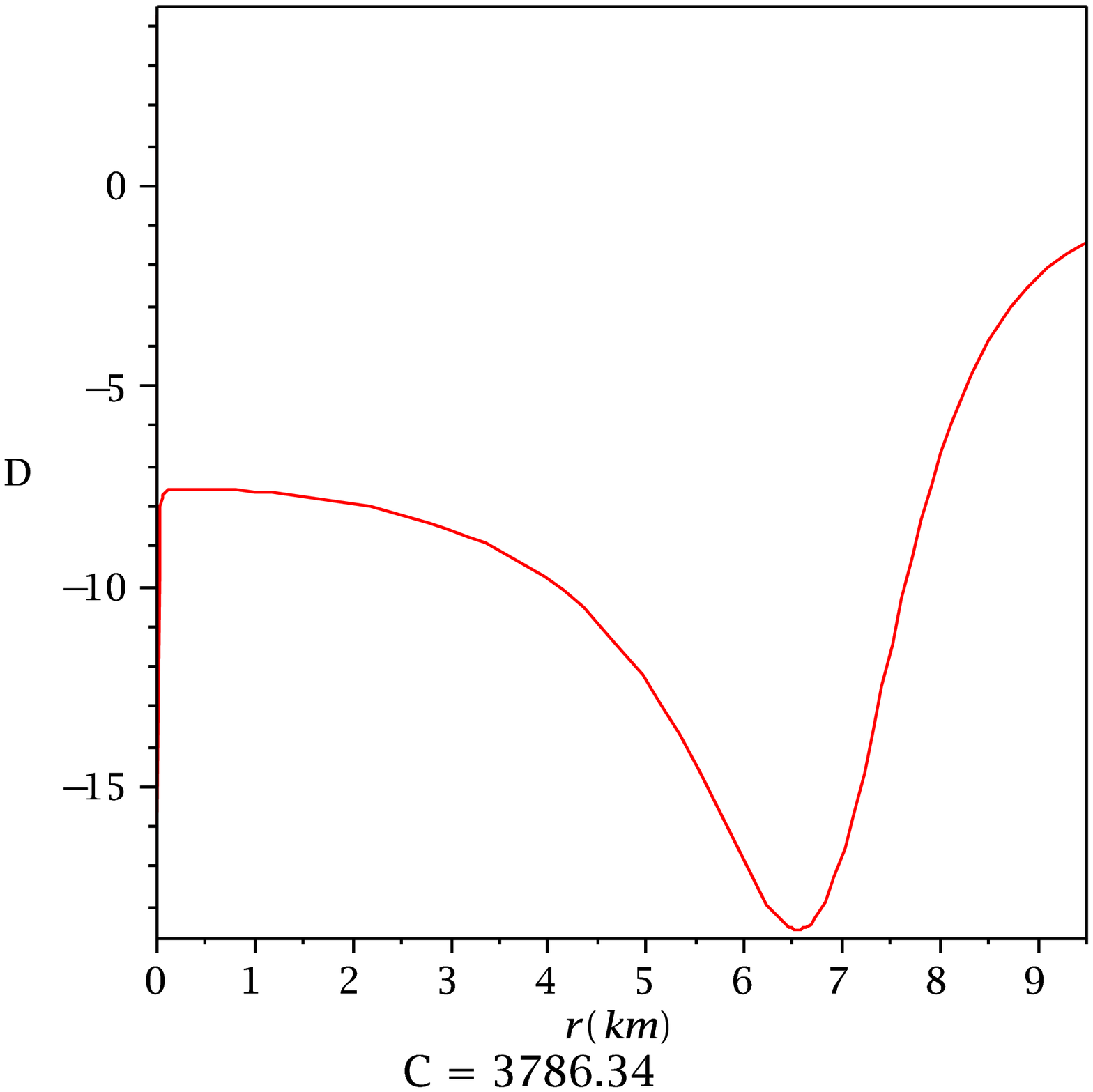}~~~~~~~
\includegraphics[height=1.5in, width=1.5in]{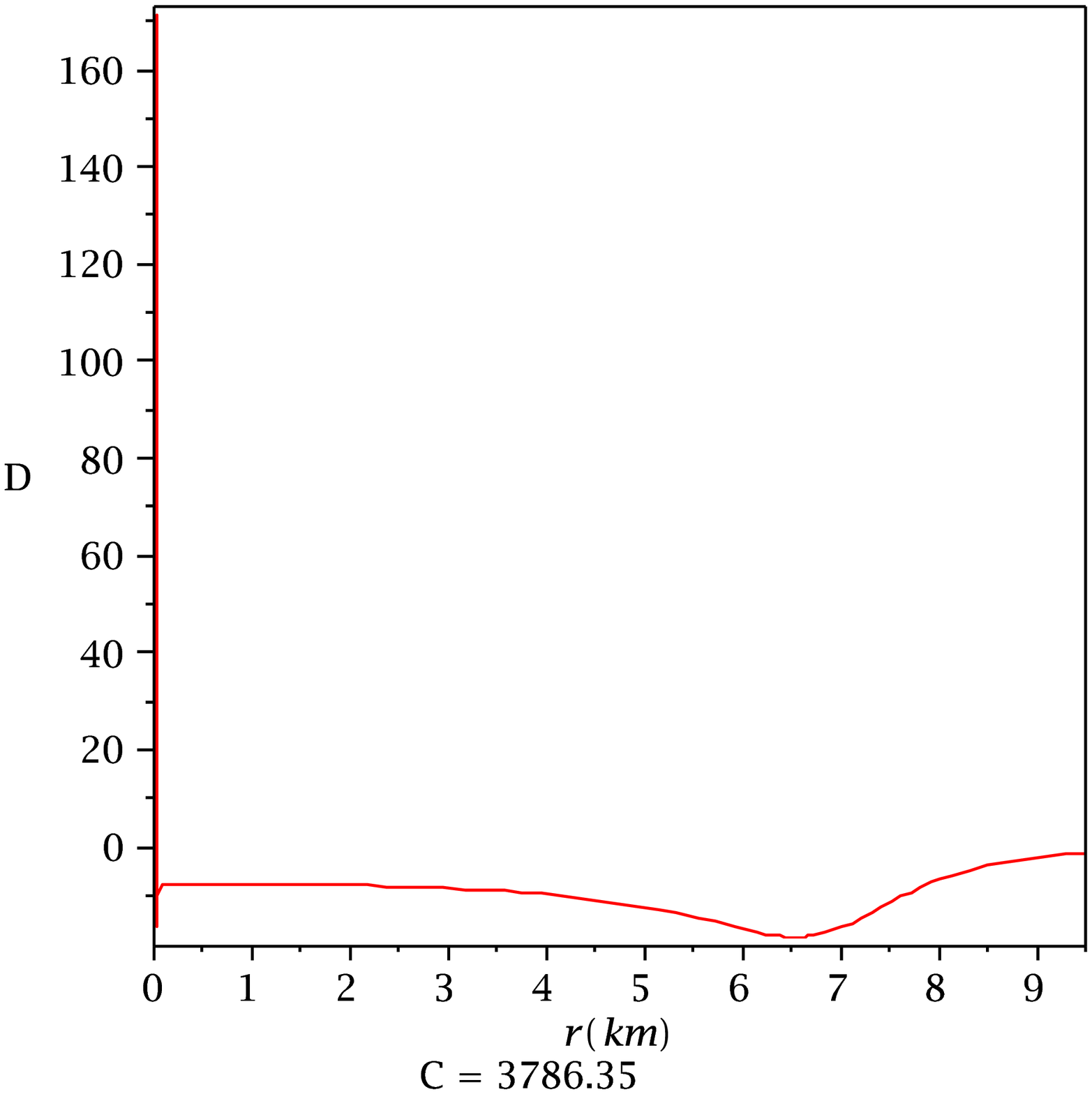}\\
FIG. 8 : Variation of $D(r)$ for different C with A=15 Km, R=9.0 Km and b (radius of the star) = 9.5 Km respectively.
\end{figure}





 Now it will be of the form $p = x\rho$ if and only if either $E = 0$ or $D = -1$. But E cannot be equal to zero for a compact star, as radius of the star is always finite. From FIG. 8 it is clear that $D = -1$  may happen according to our model if the value of C in Tolman-VII metric is  $< 3786.34$  when we take A = 15 and radious of the star is $\leq 9.5$. So, for this type of compact star of radious less than 10, EOS may be of the form $p=x\rho - T$ for non zero T, if $ C \leq 3786.34$. Otherwise it must be of the form $p = x \rho$. Surprisingly, from the FIG.8, we observe that for a star of radius upto 7 km (approximately) the matter behaves like phantom! Therefore it demands further investigation.

\section{Discussion}

In this article, we have investigated the nature of the compact
stars in the low-mass X-ray binaries(LMXBs) namely KS 1731-260, EXO 1745-248 and 4U 1608-52\citep{Guver2010,Ozel2012a,Ozel2009b} by considering it as isotropic in nature (which Tolman \citep{Tolman1939} also assumed) and the space-time
of it to be described by Tolman VII metric.\\

The results are quite interesting, which are as follows: (i) Density and pressure variation of the interior of the compact stars in LMXBs  namely KS 1731-260, EXO 1745-248 and 4U 1608-52 are well behaved[FIG.1 and FIG.2].(ii) It satisfies TOV equation and energy conditions [FIG.1,FIG.2 and FIG.3].(iii) Our compact star model is well stable according to Herrera
stability condition \citep{Herrera1992}. (iv) From mass-radius relation, any interior features of the compact stars in LMXBs namely KS 1731-260, EXO 1745-248 and 4U 1608-52 can be evaluated.(v) According to our model the surface redshift for the compact stars in LMXBs namely KS 1731-260, EXO 1745-248 and 4U 1608-52  of radius $10.5, 9.5, 9.8$ km are found to be $0.7029, 0.3744, 0.4065$  and also in the absence of the cosmological constant we get $z_{s}\leq 2$ which is
satisfactory\citep{Buchdahl1959}.\\
Therefore overall observation of the physical behavior of compact stars in LMXBs namely KS 1731-260, EXO 1745-248 and 4U 1608-52  under Tolman VII metric is justified.
\section*{Acknowledgment} MK and SMH gratefully acknowledge support
 from IUCAA, Pune, India under Visiting Associateship under which a part
  of this work was carried out.

\end{document}